\documentclass[aps,prd,twocolumn,groupedaddress, showkeys, showpacs]{revtex4}
\usepackage{graphicx,subfigure,enumerate,pstricks,latexsym,longtable,amsfonts,amsmath}
\usepackage[english]{babel}

\def\prd{Phys. Rev. D}

\def\mnras{Monthly Notices of the Royal Astronomical Society}
\def\apj{The Astrophysical Journal}
\def\aap{Astronomy and Astrophysics}

\def\apjl{Astrophysical Journal Letters}

\def\nat{Nature}

\def\araa{Annual Review of Astronomy and Astrophysics}

\def\prd{Phys. Rev. D}

\def\mnras{Mon. Not. R. Astron Soc.}
\def\apj{The Astrophysical Journal}
\def\aap{Astron. Astrophys.}

\def\apjl{Astrophysical Journal Letters}

\def\nat{Nature}

\def\araa{Annu. Rev. Astron. Astrophys.}

\usepackage{tabularx}
\newcommand{\Schw}{Schwarzschild}

\newcommand{\beq}{\begin{equation}}
\newcommand{\eeq}{\end{equation}}
\newcommand{\bea}{\begin{eqnarray}}
\newcommand{\eea}{\end{eqnarray}}

\newcommand{\ce}{{\cal{E}}} \newcommand{\cl}{{\cal{L}}} \newcommand{\cb}{{\cal{B}}}

\def\p{p}  \def\cp{\pi}
\def\x{x} 

\def\mJ{{\rm I}} 
\def\mD{{\rm II}} 

\def\af{\zeta}

\providecommand{\dif}{\mathrm{d}} \def\d{\dif}

\graphicspath{{pic/}}

\begin{document}

\title{Determination of chaotic behaviour in time series generated by charged particle motion around magnetized Schwarzschild black holes}

\author{
Radim P{\'a}nis, Martin Kolo\v{s} and Zden\v{e}k Stuchl{\'i}k
}
\affiliation{Institute of Physics and Research Centre of Theoretical Physics and Astrophysics, Faculty of Philosophy and Science, Silesian University in Opava, \\
Bezru{\v c}ovo n{\'a}m.13, CZ-74601 Opava, Czech Republic}

\begin{abstract}

We study behaviour of ionized region of a Keplerian disk orbiting a Schwarzschild black hole immersed in an asymptotically uniform magnetic field. In dependence on the magnetic parameter $\cb$, and inclination angle $\theta$ of the disk plane with respect to the magnetic field direction, the charged particles of the ionized disk can enter three regimes: a) regular oscillatory motion, b) destruction due to capture by the magnetized black hole, c) chaotic regime of the motion. In order to study transition between the regular and chaotic type of the charged particle motion, we generate time series of the solution of equations of motion under various conditions, and study them by non-linear (box counting, correlation dimension, Lyapunov exponent, recurrence analysis, machine learning) methods of chaos determination. We demonstrate that the machine learning method appears to be the most efficient in determining the chaotic region of the $\theta-r$ space. We show that the chaotic character of the ionized particle motion increases with the inclination angle.  For the inclination angles $\theta \sim 0$ whole the ionized internal part of the Keplerian disk is captured by the black hole.

\end{abstract}

\date{\today}

\keywords{black holes, magnetic fields, charged particle motion, time series, chaos and regularity, Lyapunov exponent, correlation dimension, machine learning}
 
\pacs{04.70.Bw, 95.85.Sz}

\maketitle

\tableofcontents

\section{Introduction}


It is well known that the test particle motion in the field of Kerr black holes is fully regular \citep{Car:1968:CMaPh:}. The same is true for both charged and uncharged particles moving in the field of Kerr-Newman black holes with an internal electromagnetic field related to their electric
or tidal charge \citep{Mis-Tho-Whe:1973:Gravitation:,Bic-Stu-Bal:1989:BAC:,Bla-Stu:2016:PYSR4:..94h6006B}. The regularity holds also for \mbox{Kerr-Newman-de Sitter} black holes \citep{Car:1973:BlaHol:}, even if they are dyonic, i.e., carrying both electric and magnetic charges \citep{Stu:1983:BULAI:}. The situation changes radically for black holes immersed in an external electromagnetic field since the test charged particle motion takes generally chaotic character \cite{1992GReGr..24..729K,Stu-Kol:2016:EPJC:}.

Most of the observed black hole candidates have an accretion disk constituted from conducting plasma which dynamics can generate a magnetic field external to the black hole. Another possibility is represented by an external galactic magnetic field that can be amplified by the black hole strong gravity. Such magnetic fields satisfy the test field approximation condition, having thus negligible effect on the spacetime structure and the motion of neutral particles, however, for particle with large specific charge the electromagnetic Lorentz force is relevant and leads generally to the chaotic motion of charged particles \cite{Fro-Sho:2010:PHYSR4:,Kov-Stu-Kar:2008:CLAQG:,Kol-Tur-Stu:2017:EPJC:}.

The exact shape and structure of the magnetic fields around compact objects is still under examination, but the uniform magnetic field assumption introduced by Wald \cite{Wal:1974:PHYSR4:} can be used as first simple approximation to more complex fields. The charged test particle motion in such an asymptotically uniform configuration has been already studied in a large variety of papers that give significant insight into the astrophysical processes in vicinity of magnetized black holes \cite{Pra-Vis:1978:Pra:,Prasanna:1980:RDNC:, Ali-Gal:1981:GRG:, Fro-Sho:2010:PHYSR4:, Kov-Kop-Kar-Stu:2010:CLAQG:, Kop-etal:2010:APJ:, Abd-etal:2013:PHYSR4:, Zah-etal:2013:PHYSR4:, 2013PhRvD..87f4042A, Kop-Kar:2014:APJ:, Shi-Kim-Chi:2014:PHYSR4:, Kol-Stu-Tur:2015:CLAQG:, Tur-Stu-Kol:2016:PHYSR4:, Stu-Kol:2016:EPJC:, Kop-Kar:APJ:2018:, 2019PhRvD..99d4012B}. In the present paper, we examine chaotic charged test particle dynamics around a \Schw{} black hole immersed in an external uniform magnetic field, resulting under special initial condition of ionized Keplerian accretion disk. 

The matter forming an electrically neutral (Keplerian) accretion disk orbiting such a magnetized black hole can get ionized (e.g. by an irradiation), and start to feel the external magnetic field. Under the influence of the magnetic field, the original purely circular motion of the electrically neutral matter has to be transformed into one of the following regimes of the motion of created charged particles \citep{Stu-Kol:2019:in_preparation:}: a) regular oscillatory motion possibly reflecting the high-frequency X-ray quasiperiodic oscillations observed in microquasars \cite{Kol-Tur-Stu:2017:EPJC:}, b) destruction of the ionized region of the disk due to the radial infall into the black hole, c) chaotic motion governing transformation of the Keplerian disk into thick toroidal structure, in combination with creation of winds (or relativistic jets for the case of rotating black holes \cite{Stu-Kol:2016:EPJC:}). 

%

We explore charged particle destiny for a variety of initial conditions, namely the electromagnetic interaction intensity parameter reflecting intensity of the magnetic field and the specific charge of the particle $\cb$, initial position of the particle given by the radius $r$, and the inclination angle of the disk to the magnetic field lines $\theta$. We concentrate our attention on the transition between the regular and chaotic character of the motion of the resulting charged particle dynamics in dependence on the initial conditions, namely for fixed values of the electromagnetic interaction parameter $\cb$, we determine the distribution of the regular and chaotic states in the plane $\theta-r$ of the initial conditions; the measure of the chaos is also reflected. 

The equations of charged particle motion around magnetized black holes are characterized generally as a system
demonstrating deterministic chaos -- its determination is not trivial and demands application of non-linear methods as the box-counting on one side, or the machine learning on the other. 

Observations/measurements of a (physical or general) quantity in time produce sequences of numbers, and hence time sequence (series) analysis is important tool in almost every corner of current science. Non-linear systems can produce time series with very complex behaviour that are on the first sight similar to the series of random numbers. Non-linear deterministic systems can demonstrate chaotic behaviour that is only apparently random behaviour. The deterministic chaos is hard to distinguish from random noise in observed data, especially when the degree of freedom of the generating non-linear system is high. 

Very well established linear tools, like Fourier analysis, can easily distinguish between regular and random sequences of numbers, but they fail to distinguish between a deterministic-chaos sequence and a random sequence, giving flat (constant) power spectral density in both cases. An observer using the Fourier analysis for analysis of chaotic data could make wrong conclusion and claim the data are completely random with no information inside. If we would like to extract any information from chaotic time series, the non-linear tools of detection should be used. In the present paper, we test behaviour of real-number sequences using non-linear tools -- namely, the box-counting, correlation dimension, Lyapunov exponent, recurrence quantification analysis (RQA) and machine learning. These non-linear tools are shortly presented in Appendix, and tested using the simple case of toy model based on the simple logistic map for regular/chaotic sequence generation. All codes of the methods presented in this article are written in the Wolfram {\it Mathematica} 11, since the {\it Mathematica} software provides the high-level programming language with already implemented subroutines (functions) for machine learning. The sequences of the tested regular/chaotic data are generated by the solution of the motion equations of the charged particle motion with various initial conditions. 

Combination of deterministic chaos and random noise in real measured data, i.e. small signal to noise ratio, can bring new problems to the task of distinguishing between chaos/noise. In our study such a problem is avoided, as we test the non-linear methods of detecting the deterministic chaos on time series of regular/chaotic data generated theoretically by the solution of the equations of the charged particle motion. 

The present paper can be considered as a preliminary study where we test the non-linear chaos detecting methods on system which can be controlled and where the dynamics is known. In future we plan to use such methods to study the timing signals (photon number time dependence) related to the quasiperiodic oscillations observed in the microquasar sources \cite{Rem-McCli:2006:ARAA:}. The non-linear test of the observed microquasar timing data has been already applied in different context \cite{Kar-Dut-Muk:APJ:2010:, Man-Gup-Cha:APJ:2016:, Suk-Grz-Jan:AAP:2016:, Hup-etal:MNRAS:2017:}.

Throughout the paper, we use the spacetime signature $(-,+,+,+)$, and the system of geometric units in which $G = 1 = c$. Greek indices are taken to run from 0 to 3.


\begin{figure*}
\includegraphics[width=\hsize]{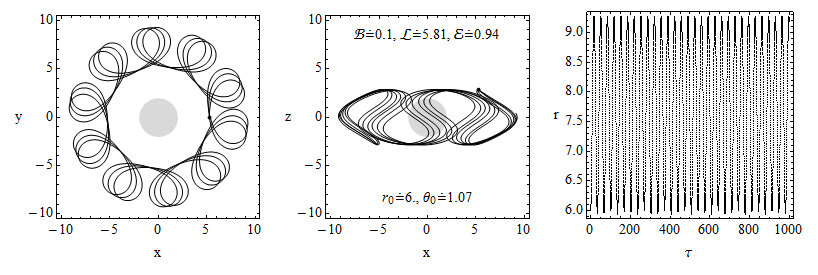}
\includegraphics[width=\hsize]{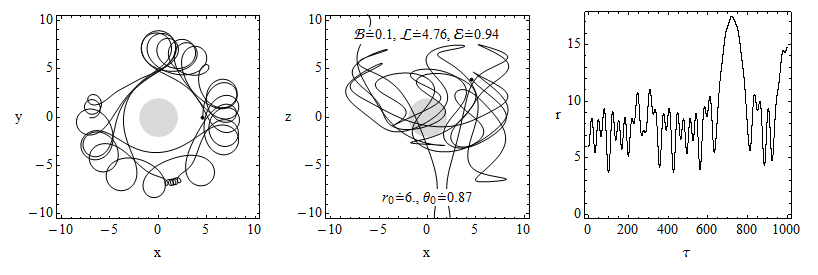}
\caption{Examples of regular (upper row) and chaotic (lower row) charged particle trajectories in the field of a magnetized \Schw{} black hole. On the left and central figures in the row, the trajectories has been plotted, while the figures on the right represent the time series of the radial coordinate $r(\tau)$. The non-linear chaos detecting methods introduced in the Appendix have been applied on such sequences of the length $10^4$, and the results are plotted in the Figures \ref{onefig}, \ref{figComparison} and \ref{figMag}. 
}
\end{figure*}

\section{Charged particle motion around magnetized Schwarzschild black holes}

\begin{figure}
\includegraphics[width=0.9\hsize]{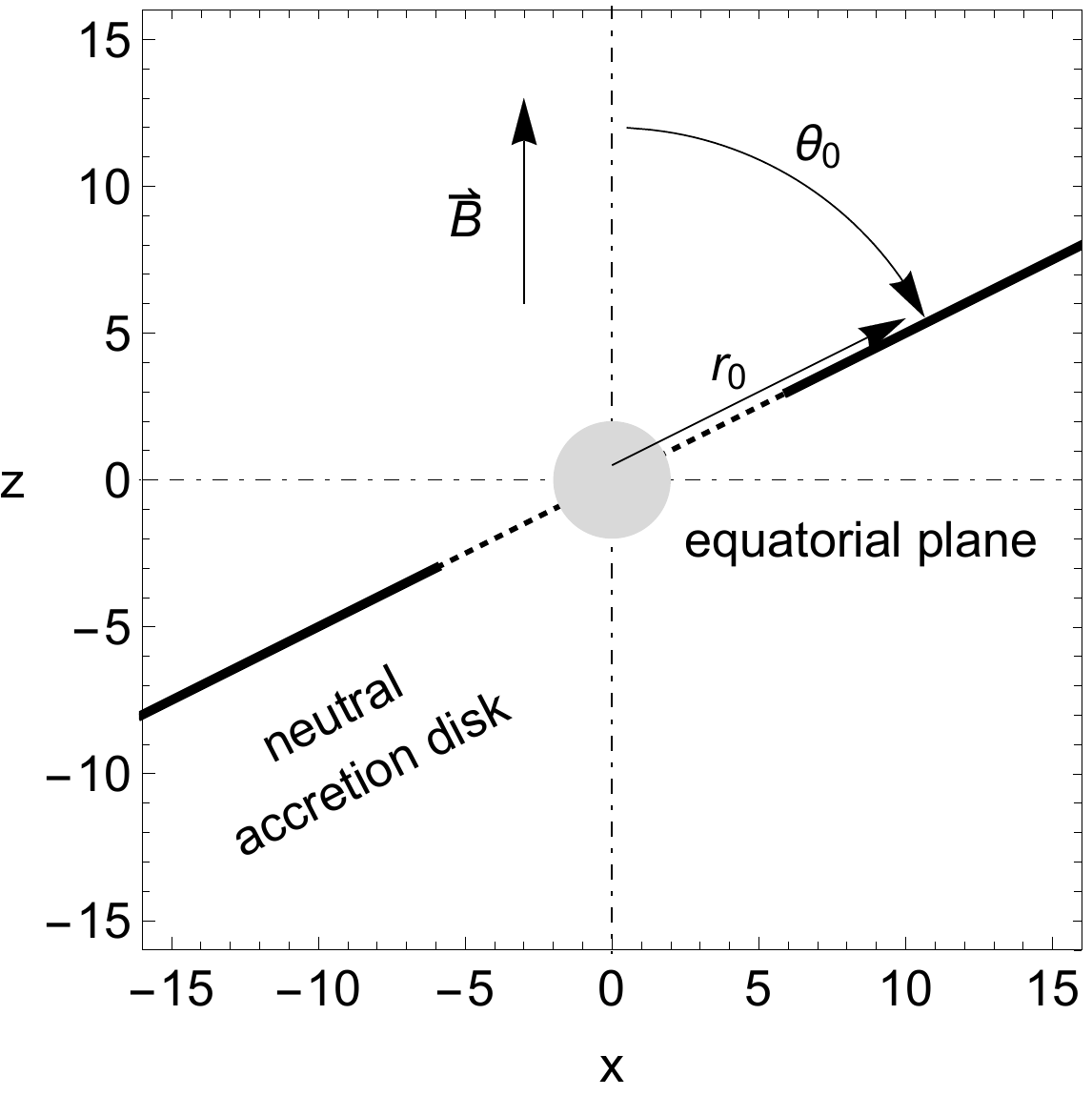}
\caption{ \label{fig01}
Schematic figure of accretion disk around black hole immersed into  
uniform magnetic field with magnetic field vector $\vec{B}$ aligned  
with $z$ axis. Neutral accretion disk plane is inclined to the $z$  
axis by $\theta_0\in(0,\pi/2)$ angle and disk inclination to  
equatorial plane is $\pi/2-\theta_0$. The accretion disk is created by  
neutral particles orbiting the black hole along circular orbits with various radii $r_0$.
For spherically symmetric Schwarzschild BHs, the presented set up of  
aligned magnetic field with $z$ axis and disk inclined to equatorial  
plane is equivalent to the disk in the equatorial plane with the magnetic  
field being inclined to the $z$ axis }
\end{figure}

The line element of the \Schw{} black hole spacetime with mass $M$ reads 
\beq
    \d s^2 = -f(r) \d t^2 + f^{-1}(r) \d r^2 + r^2(\d \theta^2 + \sin^2\theta \d \phi^2), \label{SCHmetric}
\eeq
where the so called lapse function $f(r)$ takes the form 
\beq 
		f(r) = 1 - \frac{2 M}{r}. 
\eeq
Hereafter, we put for simplicity $M=1$, i.e. we use dimensionless radial coordinate $r$ (and time coordinate $t$).

We assume an asymptotically uniform magnetic field having strength $B$ at the spatial infinity (i.e. at large distances from the black hole). The magnetic field lines are oriented perpendicularly to the equatorial plane of the black hole spacetime. The only non-zero covariant component of the potential of the electromagnetic field takes the form \cite{Wald:1984:book:}
\beq
A_{\phi} = \frac{B}{2} \, g_{\phi\phi}  = \frac{B}{2} \, r^2 \sin^2 \theta. \label{aasbx}
\eeq

The Hamiltonian governing the test (non-radiating) charged particle motion can be written in the form \cite{Wald:1984:book:}
\beq
  H_{\rm p} =  \frac{1}{2} g^{\alpha\beta} (\cp_\alpha - q A_\alpha)(\cp_\beta - q A_\beta) + \frac{1}{2} \, m^2
  \label{particleHAM},
\eeq
where the kinetic four-momentum $\p^\mu = m u^\mu$ is related to the generalized (canonical) four-momentum $\cp^\mu$ by the relation
\beq
 \cp^\mu = \p^\mu + q A^\mu, \label{particleMOM}
\eeq
that satisfy the Hamilton equations presented in the form 
\beq
 \frac{\d \x^\mu}{\d \af} \equiv \p^\mu = \frac{\partial H}{\partial \cp_\mu}, \quad
 \frac{\d \cp_\mu}{\d \af} = - \frac{\partial H}{\partial \x^\mu}. \label{Ham_eq}
\eeq
The affine parameter $\af$ is related to the proper time $\tau$ of the particle by the relation $\af=\tau/m$.

Due to the symmetries of the Schwarzschild spacetime (\ref{SCHmetric}) and the asymptotically uniform magnetic field (\ref{aasbx}), one can easily find the conserved quantities of the particle motion -- the energy and the axial angular momentum that can be expressed as 
\bea
 E &=& - \cp_t = m f(r) \frac{\d t}{\d \tau}, \label{energy} \\
 L &=& \cp_\phi = m r^2 \sin^2\theta \left(\frac{\d \phi}{\d \tau} + \frac{q B}{2m} \right). \label{angmom}
\eea

The dynamical equations for the charged particle motion in the Cartesian coordinates can be found using the coordinate transformations 
\beq
 x = r \cos(\phi) \sin(\theta),\, y = r \sin(\phi) \sin(\theta),\, z = r \cos(\theta). \label{Coord}
\eeq

Introducing for convenience the specific constants of the motion, energy $\ce$, axial angular momentum $\cl$, and the magnetic parameter $\cb$ governing the intensity of the electromagnetic interaction, by the relations \cite{Fro-Sho:2010:PHYSR4:,Kol-Stu-Tur:2015:CLAQG:}
\beq
\ce = \frac{E}{m}, \quad \cl = \frac{L}{m}, \quad \cb = \frac{q B}{2m}, \label{cbdef}
\eeq
one can rewrite the Hamiltonian (\ref{particleHAM}) in the form
\beq 
H = \frac{1}{2} f(r) \p_r^2 + \frac{1}{2r^2} \p_\theta^2  + \frac{1}{2} \frac{m^2}{f(r)} \left[ V_{\rm eff}(r,\theta) - \ce^2 \right], \label{HamHam} 
\eeq
where $V_{\rm eff}(r,\theta; \cl,\cb)$ denotes the effective potential governing the turning points of the radial and latitudinal motion, given by the relation 
\beq 
V_{\rm eff} (r,\theta) \equiv f(r) \left[1+\left(\frac{\cl}{r \sin{\theta} } - \cb\, r \sin{\theta}\right)^2\right]. \label{VeffCharged} 
\eeq
The terms in the parentheses corresponds to the central force potential given by the specific angular momentum $\cl$, and electromagnetic potential energy specified by the magnetic parameter $\cb$.

The effective potential (\ref{VeffCharged}) demonstrates clear symmetry $(\cl,\cb)\leftrightarrow(-\cl,-\cb)$ that allows to distinguish in the following only two situations corresponding to the {\it minus configuration} ($\cl>0, \cb<0$) where the magnetic parameter and the axial angular momentum have opposite signs and the Lorentz force is attracting the charged particle to the $z$-axis towards the black hole, and to the {\it plus configuration} ($\cl>0, \cb>0$) where magnetic parameter and the axial angular momentum have the same signs and the Lorentz force is repulsive, acting outward the black hole. The positive angular momentum of a particle, $\cl>0$, means that the particle is revolving in the counter-clockwise motion around the black hole. If charge of the particle is considered to be positive, $q>0$, the minus configuration, $\cb<0$, corresponds to the vector of the magnetic field $\vec{B}$ pointing downwards, while plus configuration, $\cb>0$, corresponds to the vector of the magnetic field $\vec{B}$ pointing upwards the $z$-axis \cite{Kol-Stu-Tur:2015:CLAQG:}.

The charged particle motion is limited by the energetic boundaries given by
\beq
 \ce^2 = V_{\rm eff} (r,\theta; \cl,\cb). \label{MotLim}
\eeq
The axial symmetry of the background of the combined gravitational and magnetic fields implies independence of the effective potential $V_{\rm eff}$ from the coordinate $\phi$ which allows us to examine $V_{\rm eff}(r,\theta)$ as a 2D function of the spherical coordinates $r,\theta$, or Cartesian $x,z$ coordinates (\ref{Coord}). The effective potential is positive outside the black hole horizon, and diverges at the horizon $r=2$. The region within the horizon is excluded from our investigation.

The effective potential (\ref{VeffCharged}) enables us to demonstrate general properties of the charged particle dynamics and has already been explored in detail in \cite{Kol-Stu-Tur:2015:CLAQG:} -- here we are directly applying these results. 

\begin{figure*}
\includegraphics[width=\hsize]{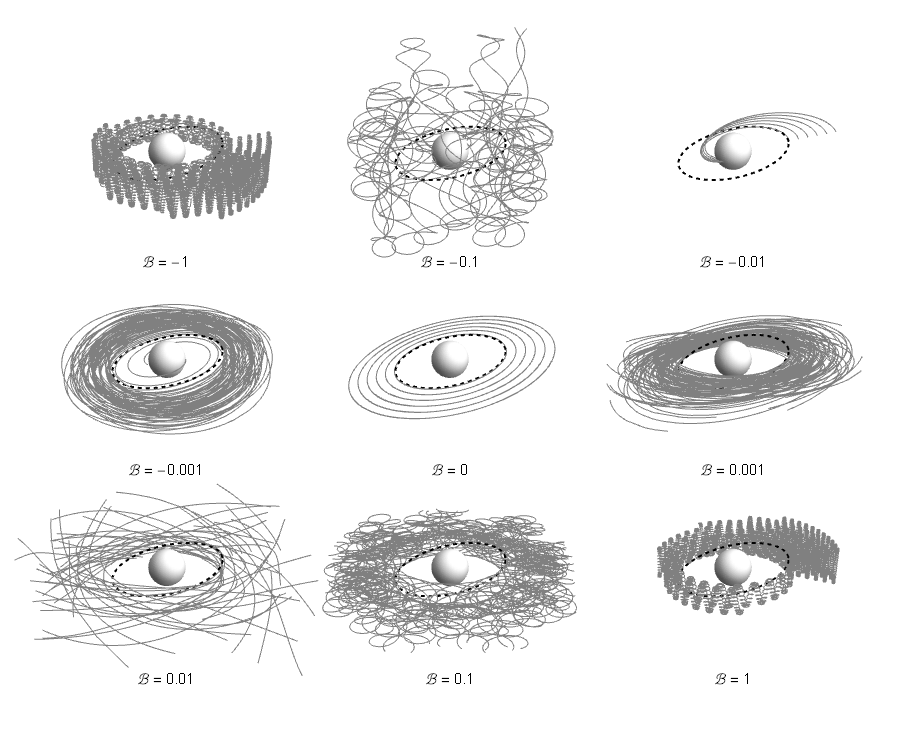}
\caption{ \label{fig_disk}
Thin Keplerian accretion disk, created by neutral test particles following circular geodesics, and its evolution (thickening/destruction) when the influence of the magnetic field is switched on. The accretion disk particles following initially circular orbits is located in the central plane with inclination $\theta_0\doteq~1.37$ to the equatorial plane, while the magnetic field lines are everywhere aligned with the $z$ axis (vertical direction). The uncharged test particles following the innermost stable circular geodesic (ISCO -- depicted by the dashed circle) of the Schwarzschild spacetime represent the inner edge of the Keplerian accretion disk. If the disk will remain neutral or the magnetic field is missing ($\cb=0$ case, middle figure) all the orbits will remain in their circular shape and we see just inclined razor thin disk. 
If a slightly strong electromagnetic interaction is switched-on ($\cb=\pm0.001$ cases, middle row), the charged particles forming the disk that is originally almost perpendicular to the magnetic field lines start to follow epicyclic oscillations around the circular orbit in both radial and latitudinal directions; the accretion disk becomes to be slightly thick. If larger magnetic field is switched-on ($\cb=\pm0.01,\pm0.1$ cases), the charged particle motion becomes quite chaotic and the accretion disk is destroyed or transformed into thick toroidal structure. The complete destruction of the Keplerian disk can be seen in the $\cb=-0.01$ case, when all the particles are captured by the black hole. If the magnetic parameter of the field that is switched-on is large ($|\cb| \geq 1$ cases), the Lorentz force dominates the particle motion. The charged particles are spiralling up and down along the magnetic field lines, while slowly moving around the black hole in the clockwise ($\cb>0$) or the counter-clockwise ($\cb<0$) direction. The thin Keplerian disk has been destroyed and transformed into some special thick toroidal structure.
}
\end{figure*}

\begin{figure*}
\includegraphics[width=\hsize]{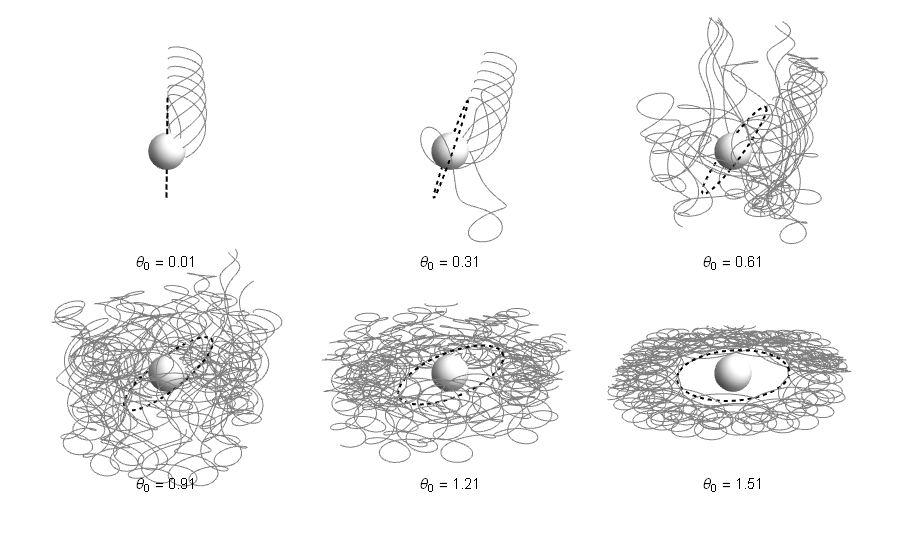}
\caption{ \label{fig_disk2}
The evolution of thin Keplerian accretion disk with various initial inclinations to the magnetic field lines is demonstrated for the magnetic parameter $\cb=0.1$. The accretion disk coexists of many circular orbits initially located in central plane given by the inclination angle $\theta_0$ related to the $z$ axis. Neutral particle ISCO represents the inner edge of the Keplerian disk and is depicted by the dashed circle. For small inclinations of magnetic field lines to the Keplerian disk, the capture by the black hole is the only option for charged particles after ionization; the accretion disk is completely destroyed in this case. For middle inclinations, the ionized particles enter quite chaotic behaviour, moving up and down along the magnetic field lines - in such case the inner region of the thin disk is transformed into a quasi-spherical structure. For Keplerian disk almost perpendicular to the magnetic field lines, the ionized particles follow regular epicyclic trajectories in vicinity of the equatorial plane, increasing slightly the disk thickness. }
\end{figure*}

\section{Ionization of the Keplerian disks}

For the magnetized \Schw{} black holes the spherical symmetry of the spacetime is combined with the axial symmetry of the uniform magnetic field. If we assume the lines of the uniform magnetic field oriented in the direction of the $z$-axis (vertical direction) \cite{Wal:1974:PHYSR4:}, the equatorial plane of the spacetime has to be perpendicular, i.e., the $x-y$ plane -- see Fig. \ref{fig01}. 

\subsection{Initial conditions of the motion of particles forming the ionized disk}

In the initial state we consider a Keplerian accretion disk orbiting the magnetized \Schw{} black hole that consists from electrically neutral test particles following circular geodesics. Due to the spherical symmetry of the \Schw{} spacetime the Keplerian disk can be located in any central plane of the spacetime, being thus inclined to the equatorial plane of the magnetized \Schw{} black hole by a latitudinal angle $\theta_0$, giving one of the initial conditions for the motion of particles of the ionized Keplerian disk, see Fig. \ref{fig01}.

\subsubsection{Ionization scenarios}

In order to demonstrate the influence of the magnetic field on an ionized Keplerian accretion disk, some realistic ionization scenarios for the neutral particles of the disk have to be considered. 
As ionization model of a neutral particle, one can consider the Magnetic Penrose Process (MPP) \cite{Par-Wag-Dad:APJ:1986:,Dadhich-etal:MNRAS:2018:}, where the original 1st neutral particle splits into two charged particles - 2nd and 3rd. Conservation of total charge of particles entering the ionization process takes the form 
\beq
               0=q_2+q_3 , 
\eeq 
and the law of conservation of the canonical momenta (\ref{particleMOM}) 
\beq
 \cp_{\alpha(1)} = \cp_{\alpha(2)} + \cp_{\alpha(3)}  \label{MPP}
\eeq
takes due to the charge conservation the form 
\beq
 p_{\alpha(1)} = p_{\alpha(2)} + p_{\alpha(3)} .  \label{MPPa}
\eeq

In many realistic scenarios, like neutron $\beta$ decay or neutral atom ionization, one of the created charged particles is much more massive then the other one, $m_{2}/m_{3}\gg~1$ -- as follows from the proton/electron or ion/electron mass ratios. The more massive charged particle (proton or ion) takes almost all the initial momentum of the original neutral particle, and the dynamical influence of the lighter charged particle (electron) can be neglected 
\beq
  p_{\alpha(1)} \approx p_{\alpha(2)} \gg p_{\alpha(3)}. \label{IonMech}
\eeq
In another realistic scenario one can consider the Keplerian accretion disk created by a plasma that can be considered as a quasi-neutral soup of charged particles - electrons and ions - orbiting around the central black hole along circular geodesic orbits. If the accretion disk is dense enough, the main free path of the charged particles will be very short in comparison to the length of the orbit around the central hole. This means that the influence of the magnetic field on the charged particle motion is effectively suppressed, and they move as a composite neutral body. However, when the density of plasma of such accretion disks significantly decreases, the charged particles start to feel effectively the influence of the magnetic field that modifies substantially their trajectories. Of course, such kind of effective ionization of the Keplerian disks is restricted to the regions located near their inner edge corresponding to the marginally stable circular geodesic (ISCO). 

 
Both previously mentioned scenarios enable a simplified ionization model, where the neutral particle just obtains an electric charge while its mechanical momentum remains conserved. Such an ionization model (\ref{IonMech}) has been already studied, but in the field of rotating Kerr black hole \cite{Stu-Kol:2016:EPJC:,Kop-Kar:APJ:2018:}, where the charged particle escape velocities and structure of escape zones were explored. Here we focus our attention on the complementary study of measure of chaos of the ionized particle motion in dependence on the particle initial conditions related to the inclination of the Keplerian disk, and the particle initial position (radius). We test which initial conditions lead to destruction of the Keplerian disk, or to transitions between the regular and chaotic motion. 

\subsubsection{Initial conditions of the charged particle motion}

We thus study the simplified particle ionization model (\ref{IonMech}) that considers only the heavy particles in the Magnetic Penrose Process, and can be characterized by simple conditions of the mass conservation and kinetic momentum conservation 
\beq
m_{(\mJ)} = m_{(\mD)}, \quad p_{(\mJ)}^{\mu} = p_{(\mD)}^{\mu}.
\eeq
The test particle kinetic momenta before (\mJ) and after (\mD) ionization are thus equal and the canonic momentum of the ionized particle is given by the relation  
\beq
\cp_{\alpha(\mD)}=p_{\alpha(\mJ)} + qA_{\alpha}. \label{ioniz}
\eeq

At the moment of ionization, the neutral or charged test particle is considered to be located on inclined circular orbit with initial position $x^\alpha$ and initial four-velocity $u_\alpha$
\bea
 x^\alpha&=&(t,r,\theta,\phi)=(0,r_0,\theta_0,0),\\
 u_\alpha&=&(u_t,u_r,u_\theta,u_\phi)=(\ce,0,0,\cl).
\eea
The specific angular momentum $\cl$ and the specific energy $\ce$ of the neutral test particles on the inclined circular orbits of the Keplerian accretion disk are given by the relations \cite{Wald:1984:book:}
\beq
 \cl_{(\mJ)} = \frac{r_0 \sin\theta_0}{\sqrt{r_0-3}}, \quad \ce_{(\mJ)} = \frac{r_0-2}{\sqrt{r_0^2 -3r_0}}. \label{CandLinSCHW}
\eeq
The simplified ionization condition (\ref{ioniz}) and use of the definitions of the energy (\ref{energy}) and the angular momentum (\ref{angmom}) enable to write the formulas of the ionized test particle specific angular momentum $\cl$ and specific energy $\ce$ in the form  
\beq
  \cl_{(\mD)} = \cl_{(\mJ)} + \cb \, r_0^2 \, \sin^2 \theta_0, \quad \ce_{(\mD)} = \ce_{(\mJ)}.
\eeq
The magnetic field has in the \Schw{} metric only one non-vanishing component, $A_\phi$ (\ref{aasbx}), hence only the (canonical) specific angular momentum $\cl$ is changed due to the ionization, while the particle specific energy $\ce$ remains constant. The energy of the neutral particles on the circular geodesic orbits, given by (\ref{CandLinSCHW}), is always $\ce\leq~1$, but energy $\ce>1$ is needed for the charged particle to escape to infinity along the $z$-axis. Therefore, no escape to infinity of ionized matter from the Keplerian disks is possible in the \Schw{} spacetime \cite{Stu-Kol:2016:EPJC:}. 

\section{Motion of charged particles forming ionized disk}

Since in the \Schw{} metric the ionized particle following originally a circular geodesic cannot escape to infinity, the capture by the black hole, or bound motion in vicinity of the original circular orbit are the only options. If the charged particle is not captured, its motion remains bounded in some closed region around the black hole -- the motion of such a bounded charged particle is in general chaotic. The character of the bounded motion depends mainly on the inclination of the original Keplerian disk, and on the magnitude of the electromagnetic interaction. As we shall see, for large inclination angles of the Keplerian disks to the magnetic field lines, $\theta_0\sim\pi/2$, the bounded motion could be regular. 

Our study of the simplified ionization of the Keplerian disks is related to the heavier charged particles resulting from the considered scenarios of the MPP process, i.e., to the motion of protons and ions, not to the electron motion in general. However, we have to note that in some special situations our results could be applied with reasonable precision also for the electron motion -- these situations have to correspond to the case when the electron has initially the specific energy and specific angular momentum close to those of the circular geodesic motion. Such a condition can be clearly satisfied for some ionized atoms, or while the electron following the circular geodesics starts to feel the influence of the magnetic field due to decreasing density of the disk. 

\subsection{Typical trajectories of charged particles: chaos and regularity}

Due to the spherical symmetry of the \Schw{} spacetimes, the Keplerian disk can be located in any central plane, contrary to the case of the Kerr black hole spacetimes where the inner parts of the Keplerian disks have to be located in the equatorial plane of the spacetime due to the so called Bardeen-Peterson effect reflecting the interplay of the disk viscous stresses and the frame dragging of the spacetime \cite{Bar-Pet:1975:ApJ:}. We thus first discuss properties of the ionized disks orbiting the magnetized \Schw{} black hole with large enough initial inclination angle, $\theta_0\sim\pi/2$, in dependence on the magnetic parameter characterizing the intensity of the electromagnetic interaction. As a second case we discuss the role of the inclination angle for the case of the magnetic interaction parameter giving the strongest chaotic behaviour for the near-equatorial disks. Finally, we study the combined role of the magnetic parameter $\cb$ and the inclination angle for charged particles located initially at the ISCO. 
 
The influence of the magnitude of the magnetic field (parameter) on the character of the ionized near-equatorial accretion disk is demonstrated in Fig. \ref{fig_disk}. When the magnetic field is missing ($\cb=0$ case), all the orbits of the ionized Keplerian disk keep their circular shape and we see just inclined razor thin disk. 
When a slight magnetic field is switched-on ($\cb=\pm0.001$ cases), the charged particles forming the ionized disk become to be unsettled, and the circular orbits are perturbed, but the ionized particle orbits are still bounded in vicinity of the equatorial plane, following oscillatory motion in the radial and latitudinal coordinates, and the accretion disk becomes slightly thick. Notice that in the case $\cb=-0.001$ the particles starting very close to the ISCO are captured by the black hole, while such trapping is not possible for the case $\cb=0.001$. 
When a larger magnetic field is switched-on ($\cb=\pm0.01$ cases), the charged particle trajectories become chaotic and a lot of trajectories finishes inside the black hole. The complete destruction of the ionized Keplerian disk can be seen for the $\cb=-0.01$ case, when all the orbits are captured by black hole. 
For even larger magnetic field ($\cb=\pm0.1$ cases), the Lorentz force becomes more influential and charged particle trajectories start to be strongly chaotic creating a toroidal structure, and demonstrate tendency to wind up and down along the magnetic field lines, especially in the case of $\cb=-0.1$ when the structure of orbiting particles resembles a sphere. 
When the magnetic field parameter is large ($\cb=\pm1$ cases), the Lorentz force becomes to be the dominant force for the particle motion. The charged particles wind up and down along magnetic field lines, demonstrating simultaneously slow shifting around the black hole in the clockwise ($\cb>0$) or counter-clockwise ($\cb<0$) direction. Once again, the thin Keplerian disk has been transformed into some thick structure spread around the equatorial plane of the magnetized black hole. For larger values of the magnetic parameter, $|\cb|>1$, similar behaviour is demonstrated by the ionized Keplerian disk.

The influence of the initial inclination angle $\theta_0$ of the Keplerian accretion disk to the magnetic field lines on the character of the motion of particles of the ionized disk is demonstrated in Fig. \ref{fig_disk2} for the magnetic parameter fixed to the value of $\cb=0.1$ corresponding to the case when the ionized near-equatorial Keplerian disks demonstrate the most chaotic behaviour. Our goal is to demonstrate clearly how the modifications of the inclination angle influence the chaotic character of the particle motion of the ionized disk. 

Clearly, for large inclinations angle of a near-equatorial Keplerian disk, $\theta_0=1.51$ case, we can see that the charged particle trajectories remain close to the equatorial plane, demonstrating epicyclic oscillations in the radial and vertical directions -- this kind of regular motion around magnetized black holes has been discussed in detail for both Schwarzschild \cite{Kol-Stu-Tur:2015:CLAQG:} and Kerr \cite{Tur-Stu-Kol:2016:PHYSR4:} black holes and can be related to the high-frequency quasiperiodic oscillations observed in microquasars \cite{Stu-Kot-Tor:2013:ASTRA:,Kol-Tur-Stu:2017:EPJC:}. For Keplerian disks that are slightly off-equatorial, $\theta_0=1.21$ case, we observe after ionization a transition to the chaotic particle motion demonstrating slight tendency to wind up and down along the magnetic field lines -- the disk seems to be deformed into a toroidal structure. For Keplerian disks with middle inclination, $\theta_0=0.91$ case, the particles of the ionized disk demonstrate strongly chaotic motion with extension comparable in both radial and vertical dimensions, causing that the transformed disk resembles a quasi-spherical structure of orbiting charged particles. If the initial inclination of the Keplerian disk is slightly lowered, $\theta_0=0.61$ case, we observe a clear tendency of the charged particle motion to follow the direction of the magnetic field lines, but the transformed disk still resembles a quasi-spherical structure. On the other hand, for small and very small initial inclination angles of the Keplerian disk, $\theta_0=0.31, 0.01$ cases, we observe fall of the charged particles into the black hole (direct fall in the case of very small inclination angle) -- in such situations the internal region of the Keplerian disk becomes unstable after ionization, being completely destroyed and captured by the black hole. 

\begin{figure}
\includegraphics[width=\hsize]{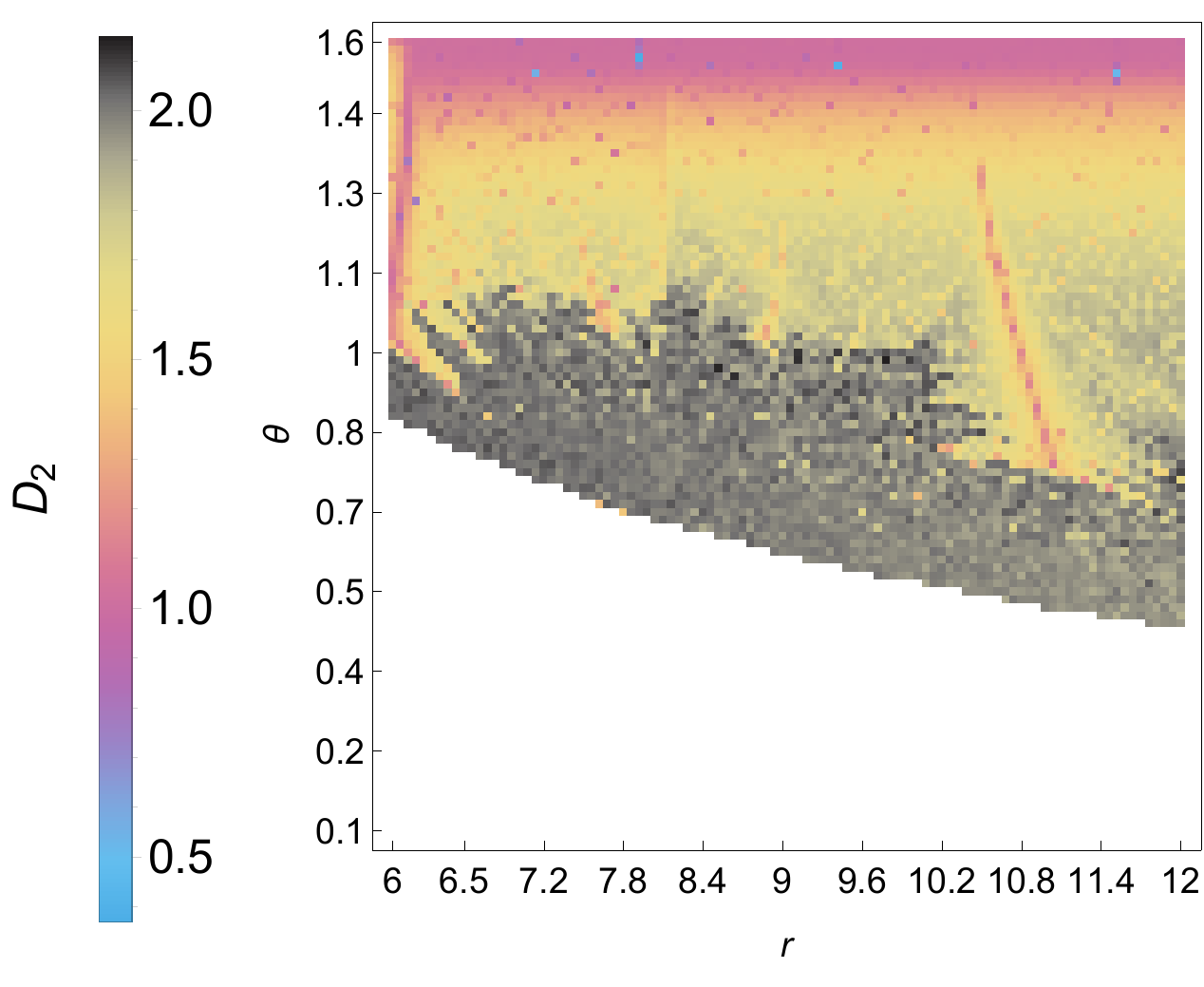}
\caption{\label{onefig} 
Determination of chaotic behaviour of ionized particle motion. The charged particle motion is represented by time series created from the time dependence of its radial component $r(\tau)$. The equi-distance sequence of points $r(\tau_i)$ is used to determine the measure of the chaos in the ionized particle motion by applying the non-linear method of Correlation dimension. Every point in this figure represents one charged particle trajectory for which the chaos-measure is determined. The ionized particles are represented by the initial disk radius $r_0$, given on the horizontal axis, and the initial disk inclination angle $\theta_0$, given on the vertical axis; the magnetic parameter has been chosen to be $\cb = 0.1$. We consider the range of disk radii $r_0\in[6, 12]$ corresponding to the inner region, and whole the range of the inclination angles $\theta_0\in[0,\pi/2]$. The chaos measure of the particle trajectory is represented by the colour of the point, white area represents the particle trajectories captured by the black hole. The colour scale related to the quantity giving the measure of chaos is shown on the bar -- increasing number means increasing chaos-measure. The gray colour thus means the upper degree of chaos. 
Since the chaos-measure quantities are defined in different ways for different non-linear methods, and their ranges are also different, in the following figures the colour mode is chosen in a way that tends to normalize the results obtained by different methods. 
In the present and all the following figures an initial Keplerian disk with given inclination angle $\theta_0$ is determined by the horizontal slice with constant inclination $\theta=\theta_0$.
}
\end{figure}

\subsection{Non-linear methods determining chaos-regularity transition}

Using various non-linear methods for determination of chaos and its measure, we realize detailed study of the stability of the Keplerian disks orbiting a magnetized Schwarzschild black hole, and the measure of chaos in the motion of particles of the ionized disk, in dependence on the initial inclination angle between the disk plane and the magnetic field lines direction $\theta_0$, and the position of the particles at the Keplerian disk $r_0$. We shall study the cases presented above, for the characteristic values of the magnetic parameter determining the intensity of the electromagnetic interaction between the ionized particles and the external magnetic field $\cb$. We first study the case of the magnetic parameter $\cb=0.1$, testing the efficiency of a large variety of the non-linear methods. The results are presented in Figs \ref{onefig}, \ref{figComparison}. We test the following non-linear methods: Box-Counting, Correlation Dimension, Lyapunov Exponent, Reccurence Quantification Analysis with variants RR (recurrence rate), DET (determinism -- predictability of the system), LL (diagonal line length), ENTR (Shannon entropy -- probability distribution of the line lengths), and Machine Learning with variants Random Forest, Neural Network, Linear Regression, Nearest Neighbours, Gradient Boosted Trees. (All of these non-linear methods of detecting chaos are presented and tested by a simple logistic map function in the Appendix.) Finally, we study in detail the cases of $\cb = \pm0.001, \pm0.01, \pm0.1, \pm1$, comparing the application of the Correlation Dimension method and the Machine Learning method with the Random Forest algorithm. 

\subsubsection{Description of the numerical methods}
\begin{figure*}
\includegraphics[width=\hsize]{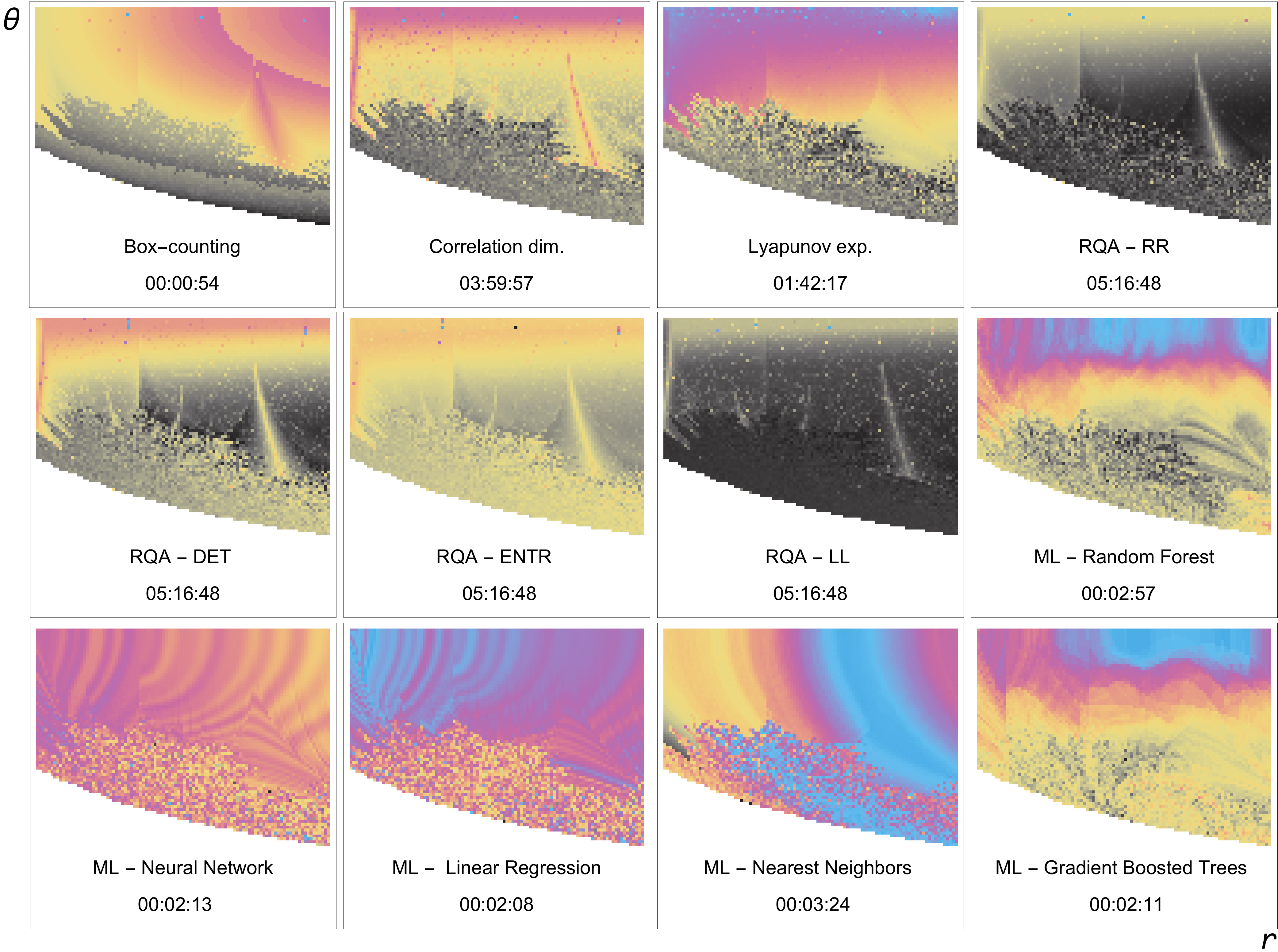}
\caption{ \label{figComparison}
Comparison of the considered non-linear methods of chaos determination and measurement, as applied on the charged particle trajectories from initially Keplerian accretion disk as described in Fig. \ref{onefig}. Every figure corresponds to the disk (initial position of ionized particle) radii $r_0\in[6, 12]$ and inclination of magnetic field lines to the Keplerian disk plane $\theta_0\in[0,\pi/2]$. Name of the used non-linear method, and total time for construction of the whole figure in format (hh/mm/ss), are also presented. The range of the quantities giving the chaos-measure in various methods are different, the colour scale is chosen accordingly, to reflect the maximum of the chaos-measure by a specific colour.
The RQA methods have been reversed by transformation. We subtracted the calculated data from the corresponding maximal data value, for RQA - LL we also have lowered one outreach maximal value to the second maximum in order to make the structure visible.
}
\end{figure*}

\begin{figure*}
\includegraphics[width=\hsize]{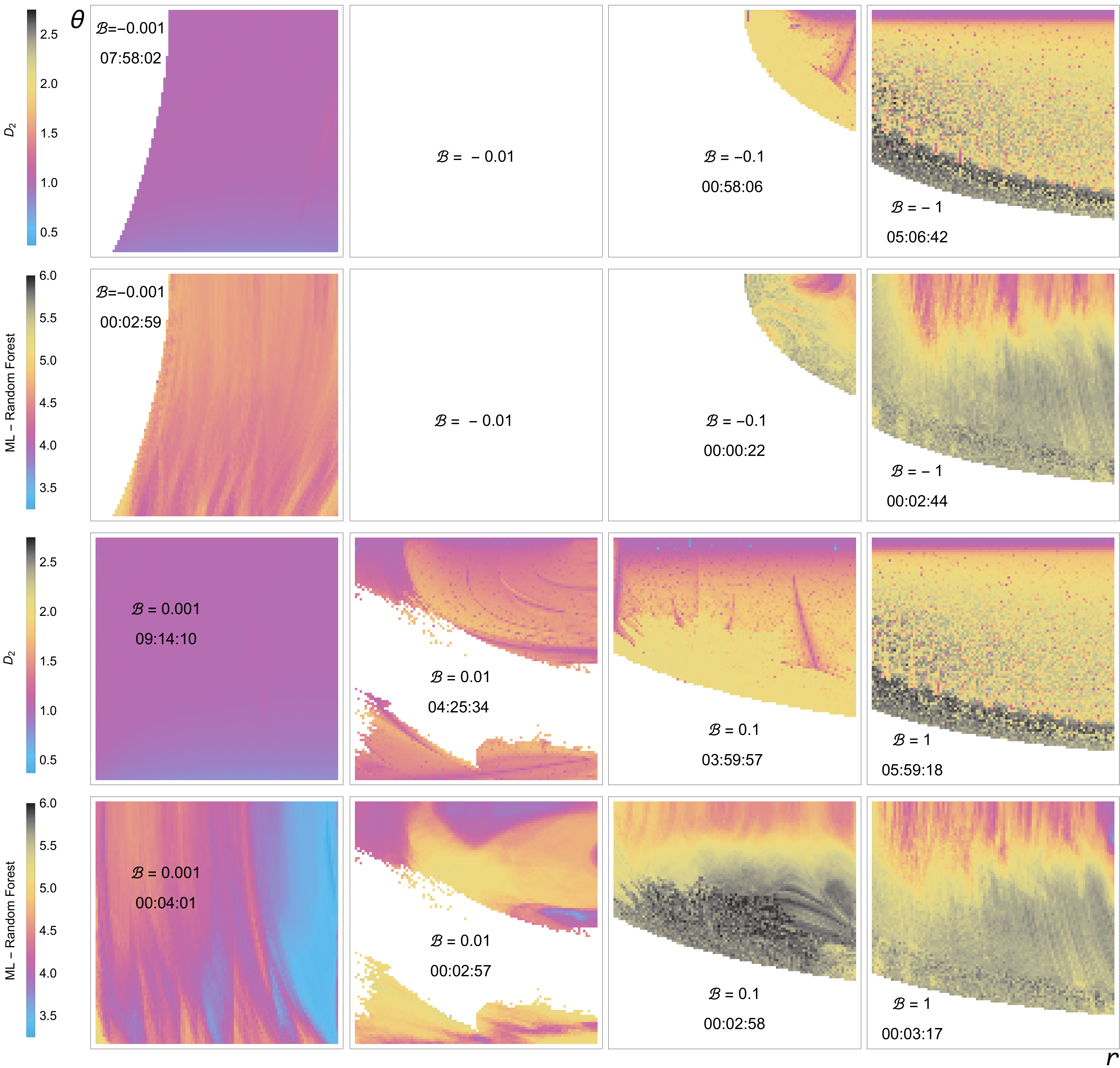}
\caption{ \label{figMag}
Evolution of an ionized Keplerian accretion disk in an external magnetic field characterized by typical magnetic parameters $\cb = \pm0.001, \pm0.01, \pm0.1, \pm1$. As in Figs \ref{onefig} and \ref{figComparison}, the Keplerian disk (initial position of the ionized particle) radius $r_0\in[6, 12]$ is given on the horizontal axis, while magnetic field inclination to disk $\theta_0\in[0,\pi/2]$ is given on the vertical axis. 
Two non-linear methods of the Correlation Dimension (upper row), and the Machine Learning with Random Forest algorithm (lower row) has been applied and their results are compared. The presented results are similar for both methods, the Machine Learning method takes considerably less amount of computer time for the calculation.
}
\end{figure*}

\begin{figure*}
\includegraphics[width=\hsize]{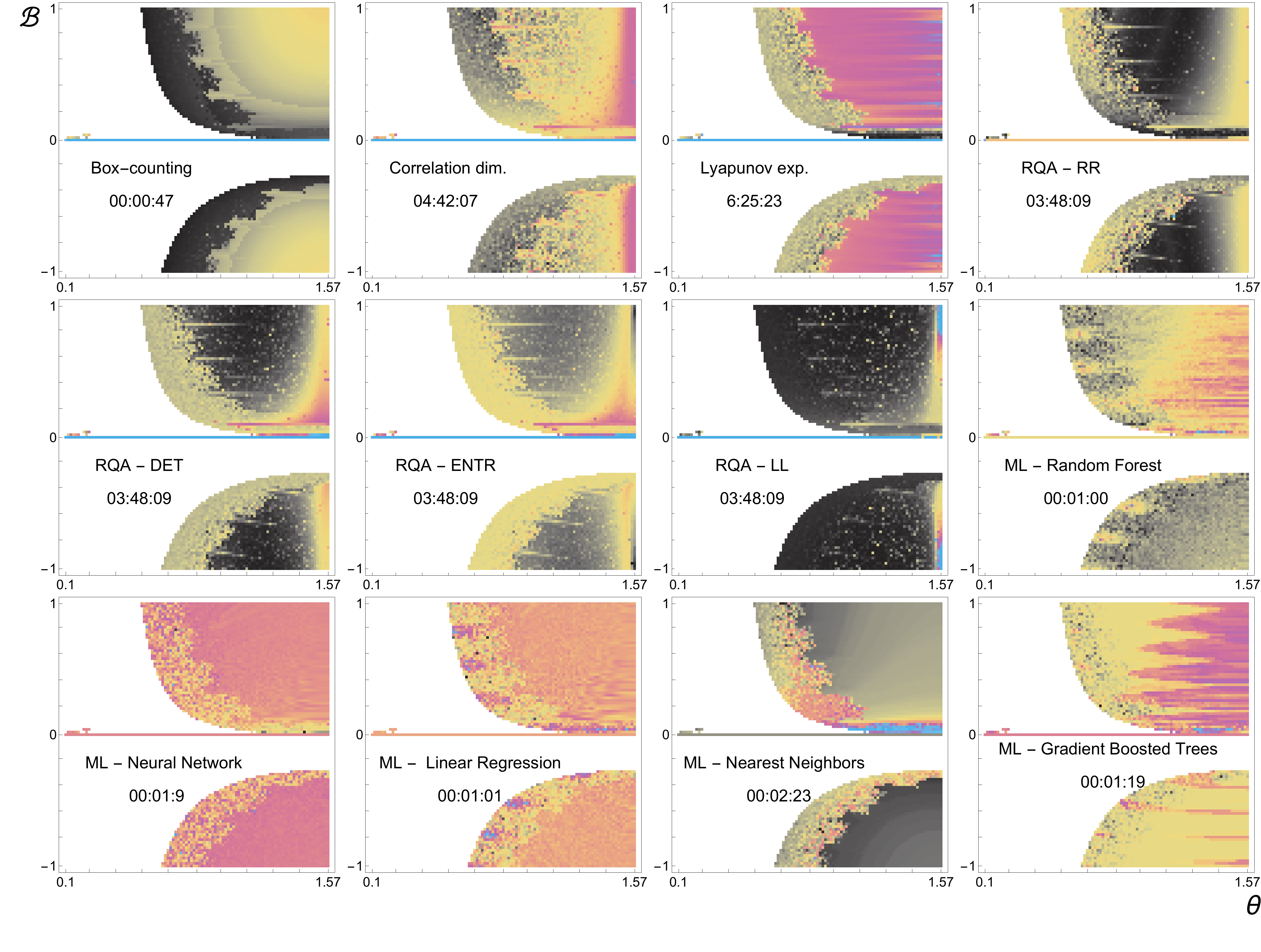}
\caption{ \label{figComparison2}
Similarly as in Fig. \ref{figComparison}, comparison of the considered non-linear methods of chaos determination and measurement is applied on the charged particle trajectories from initially Keplerian accretion disk as described in Fig. \ref{onefig}, but instead of the radial coordinate, the magnetic field parameter $\cb$ is used. Therefore every figure corresponds to the disk  magnetic field parameter $ \cb \in[-1, 1]$ and inclination of magnetic field lines to the Keplerian disk plane $\theta_0\in[0,\pi/2]$, while the initial position of ionized particle  is fixed to the ISCO value $r_0 = 6 $. For both parameters $ \cb $ and $\theta $ we used 101 equidistantly distributed values. Name of the used non-linear method, and total time for construction of the whole figure in format (hh/mm/ss), are also presented. The range of the quantities giving the chaos-measure in various methods are different, the colour scale is chosen accordingly, to reflect the maximum of the chaos-measure by a specific colour.
 As in Fig. \ref{figComparison}, the RQA methods have been reversed by transformation. We subtracted the calculated data from the corresponding maximal data value, for RQA - RR, ENTR, LL we also have lowered some outreach maximal values in order to make the structure visible.
}
\end{figure*}

For numerical trajectory calculations, the Wolfram {\it Mathematica} software was used with fixed step size fourth order explicit Runge-Kutta method. The size of the step has been set to $ 10^{-2} $ and the time of integration to $10^4$, then every 100-th point was taken; therefore, all trajectories which have not fallen into the black hole are of the length of $10^4$ points. 
The fixed step method has been used to ensure that the non-linear methods for calculation of the chaos-measure of every pair of the initial conditions take the input of the same length. All the trajectories that fall into the black hole were not taken as input for calculation of the chaos-measure, and are denoted by white colour in the final figures \ref{onefig}, \ref{figComparison}, \ref{figMag}, and \ref{figComparison2}. Initial conditions for the charged particles from the ionized Keplerian disk radius $r_0$ have been taken from interval $r_0\in[6, 12]$, while inclination between the disk plane and the magnetic field lines $\theta_0$ has been taken from interval $\theta_0\in[0,\pi/2]$. Both these intervals have been evenly distributed among $10^2$ points; all their combinations lead to $10^4$ couples of conditions for particle trajectories, plotted for every figure from Figs \ref{onefig}, \ref{figComparison}, \ref{figMag}, and \ref{figComparison2}.
All the computations have been realized on DELL 7577 with processor Intel® Core i7-7700HQ and 16GB 2400MHz DDR4 SDRAM.


As can be seen from Fig. \ref{onefig}, the ionized particle trajectories are mostly regular when the Keplerian  disk is near the equatorial plane, i.e., the inclination angle is of the value $\theta\sim\pi/2$ and the magnetic field is almost perpendicular to the disk plane. As the inclination angle is gradually decreased, we see gradual increase of the chaotic character of the charged particle motion. Near the inclination angle $\theta\sim~1.1$, we enter a fractal like zone, where the strongly chaotic trajectories are located in close vicinity to those not so much chaotic (or regular). Finally, when the inclination angle is small enough (say $\theta\sim~0.7$ for $r=7$) all the trajectories will eventually end being captured by the black hole. This "capture inclination angle" depends on the radial position inside the disk - trajectories with initial position closer to the black hole get captured for larger inclination angle. 
When the magnetic field will be parallel (almost parallel) to the ionized Keplerian disk ($\theta\sim~0$), all the trajectories will be captured by BH. But one can argue, that such configuration is not so realistic, since the currents inside the ionized accretion disk always produce some internal magnetic field with component parallel to the disk plane. 

Except the fate of the captured parts of the ionized Keplerian disks, we have to follow the fate of the rest of such disks whose charged particles remain bound in vicinity of the black hole, following generally chaotic trajectories with possible exceptions of special regular motions. The measure of the chaotic character of the motion, and existence of possible "islands of regularity" can be treated by the non-linear methods of determination of deterministic chaos. 

All the non-linear methods we have used are revealing similar structures at the Figs. \ref{figComparison}, \ref{figMag}, and \ref{figComparison2}. The colour scale (colour palette) is different for each method, because each of the non-linear methods denotes trajectories by different scale of the chaos-measure estimation -- see section \ref{section_logistic}. Therefore, the sensitivity of different non-linear methods to the chaotic behaviour is different, so it seems to be quite complex task to use one common colour scale for all methods. Moreover, when plotting various chaos estimations in one figure, the presence of outreach values is inconvenient -- if there is some value far away from the others, then it is problematic to cover the differences between closer values by visible colour differences. At Figs. \ref{onefig}, \ref{figComparison}, \ref{figMag}, and \ref{figComparison2}, we should focus on structures produced by different colours - such structures will indicate transition between regular and chaotic motion. 

Box-Counting method \ref{boxcount} (see Appendix) is clearly the fastest method and that is a big advantage while studying large datasets. The reason is the simplicity of the algorithm which does not need to make many numerical operations, and therefore does not need much computing power. We also have used only one-dimensional input in the sense of non-embedding dimension (for the Box-Counting, but for the Correlation Dimension we did use it), which non-linearly enlarges the computing, while its application did not seem to improve our final figures.
The Correlation Dimension method \ref{correlation} is, along with the Lyapunov Exponent \ref{lyapunov} method, the most time dependent, when we consider the fact that from Recurrence Quantification Analysis (RQA) method \ref{xRQAx} computation we get four figures. This property varies also with the density of points in time series. 
For the Lyapunov Exponent method, together with ML methods, the colour range seems to be highly continuous, while the Correlation Dimension method shows some structure details, similarly to the RQA methods.
RQA demonstrates similar structure for all the tools (RR - recurrence rate, DET - determinism, LL - line length, ENTR - entropy). Time needed for the RR variant could be shorter than for the DET variant, which also needs lower computing time than the ENTR and LL variants. This is caused by the necessary numerical operations needed for computing each of them -- while more complicated numerical operations for the ENTR and LL variants are already done, the RR and DET variants can be derived from them, but not vice versa. As already mentioned, sometimes in the results of these non-linear methods a big gap between the most and the least chaotic trajectory occurs, implying that the trajectories in some interval of the chaos-measure are not distinguished by distinct colours, and the chaotic structure is not observable with sufficient precision. This case can be handled by adjusting of the outreach values, or by defining special rules for plotting. 
The implementation of the ML models \ref{machinelearning} is also very fast, as the time needed to build up a model has not been counted in Figs \ref{figComparison}, \ref{figMag}, and \ref{figComparison2}. When considering time to build up such a model, several factors have to be taken into account, like choice of the programming language and its library, concrete algorithm and its parameters, data preprocessing, model validation, etc.  We can observe that different ML algorithms are producing structures which are not so similar, as in the case of the previous methods (with the same training set input). This is affected by the fact that behind each ML algorithm is different principle of achieving the goal (data classification or regression). It has been proved that some of them are relevant mostly for some specific tasks -- by this not only the specific application for particular tasks like, e.g., image recognition is meant, but also the usage according to available data, the model generalization (bias - variance trade-off), memory consumption, or the possibility of additional customization (neural network). 

\subsubsection{Evolution of ionized Keplerian accretion disk}

We have assumed simple ionization scenario for particles from a neutral Keplerian disk, where the neutral particles will split into two oppositely charged particles, and the newly charged particles start to feel influence of an inclined external uniform magnetic field. In the most common process of such ionization, the produced charged particles have quite different masses - proton for example is by three orders more massive than electron, and hence takes almost whole initial momentum of the former neutral particle, see (\ref{IonMech}). On the other hand, the electron, with specific charge $q/m$ by three orders higher than that of the proton, experiences the influence of the magnetic field due to the Lorentz force by three orders of magnitude stronger than the proton, i.e., for electrons the magnetic parameter $\cb$ will be higher by three orders in comparison with those related to protons -- see eq. (\ref{cbdef}). In order to consider all possible inclinations of the disk to the magnetic field lines, we restrict our attention to the magnetized non-rotating Schwarzschild black holes when particle energy after ionization process remains constant. 
In the present paper we have studied the motion of the heavier particles (protons, ions) after the ionization, such particles at bounded orbits around black hole will mix with neutral particles of the Keplerian disk and form new deformed lightly charged accretion disk, or the inner region of the disk could be ionized completely.
The electrons are assumed to escape from the black hole neighbourhood in the form of wind, or form a corona around the accretion disk.

In Fig. \ref{figMag} all possible inclinations of magnetic field vector to the accretion disk plane $\theta\in(0,\pi/2)$ are considered for various characteristic values of the interaction with the magnetic field $\cb$. As shown in Fig. \ref{fig_disk}, the accretion disk destiny strongly depends on the magnetic field strength.

For very small magnetic field magnitudes $|\cb|=0.001$ the motion of ionized particles is almost regular for any inclination and surprisingly it is more regular when the inclination of the magnetic field is small. Also almost all particle trajectories  will remain in disk, only the inner edges of the disk in $\cb=-0.001$ case are captured by the Schwarzschild black hole.

For larger magnetic field magnitudes, the situation is completely different and more complicated - the ionized accretion disk has completely different evolution in the $\cb>1$ and $\cb<1$ cases. For $\cb=-0.01$ (and $\cb=-0.1$) all (and almost all) ionized trajectories are captured by the black hole, and the disk (inner parts of the disk) will be destroyed. For $\cb=0.01$, only the disks with small and large inclinations are preserved, for large inclinations some thick disk of ionized particles will be formed, for small inclinations the ionized trajectories are forming some sphere around the central black hole. The case $\cb=0.1$ has been already shown in Figs. \ref{onefig} and \ref{figComparison} - disks with small inclination are destroyed, mildly inclined disks are formed by particles following strongly chaotic trajectories creating a (quasi-)spherical configuration, and charged particles from disks with almost perpendicular inclination of magnetic field will remain near the disk plane.

If the magnetic field parameter is really large, $|\cb|\geq1$, as it could be for elementary particles with large specific charge $q/m$, the Lorentz force is leading force in the system. The charged particles are winding up and down in vertical direction along the magnetic field lines, while slowly moving along the central black hole. Both cases $\cb=-1$ and $\cb=1$ are giving the same resulting trajectories, differing only in the direction of slow motion around the central black hole -- see Fig. \ref{fig_disk}.  

Our numerical study thus determines the relation between  regions of regular and chaotic motion of the charged particles of the ionized Keplerian disk. Moreover, also the critical "capture inclination angle" is determined in dependence on the initial position in the Keplerian disk for fixed values of  magnetic field parameter~$\cb$. 

In order to illustrate in detail the role of the combined effect of the magnetic field parameter $\cb$ and the disk inclination angle, we study distribution of the capture by the black hole, and the relation of the regular and chaotic motion, for charged particles located initially at the ISCO. The results are given in Fig. \ref{figComparison2} for $\cb \in[-1, 1]$ and inclination $\theta_0\in[0,\pi/2]$. All the non-linear methods of chaos determination give again the same results clearly reflecting the  asymmetry of the capturing process in relation to the sign of the magnetic parameter.

{ 

\subsection{Astrophysical relevance}

We have studied the influence of an external uniform magnetic field showing that even small $\cb$ parameter has significant influence on the thin Keplerian accretion disk, if the particles of the disk become ionized, as shown in Fig. \ref{fig_disk}. Such qualitative description should be followed by qualitative analysis to clear up if all these effect are astrophysically relevant.

In this article we assume the magnetic field to be uniform, but real magnetic fields around microquasar or supermassive black holes, and their accretion disks, are far away from being completely regular and uniform. The Wald uniform magnetic field solution is used as a useful approximation describing properly at least the magnetic field magnitude. Moreover, the magnetic parameter $\cb$ contains, together with the field strength, also the specific charge of the ionized test particles, see Eq. (\ref{cbdef}). This implies that in order to make proper estimation of the magnetic field magnitude, one needs to identify first the type of matter inside accretion disk. 

In our approach, the "charged particle" can represent matter ranging from electron to some charged inhomogeneity orbiting in the innermost region of the accretion disk. The specific charges $q/m$ for any of such structures will then range from the electron maximum to zero. Recalling the physical constants in the dimensionless magnetic parameter as $\cb = |q| B G M/(2 m c^4)$, we get from eq. (\ref{cbdef}) the magnetic field strength in Gauss
\beq 
B = \frac{2 m c^4 \cb}{q G M} ~ [{\rm G}], \label{MFgauss} 
\eeq
where the quantities are given in CGS units, see Tab. \ref{tabX}.

\begin{table}[!ht]
\begin{center}
\begin{tabular}{c@{\qquad} c@{\quad} c@{\quad} c@{\quad} c } 
\hline
 $\cb=0.01$ & electron & proton & charged dust \\
\hline
\hline
$M={10~{M}_{\odot}}$ & $2\cdot10^{-5}~{\rm{G}}$ & $ 0.04~{\rm{G}} $ &  $ 2\cdot10^8~{\rm{G}}$\\
$M={10^8~{M}_{\odot}}$ & $2\cdot10^{-12}~{\rm{G}}$ & $ 4\cdot10^{-9}~{\rm{G}} $ &  $ 20~{\rm{G}}$\\
\hline
\end{tabular}
\caption{Magnitudes of magnetic fields $B$ in Gauss units for magnetic parameter $\cb=0.01$ and stellar $M={10~{M}_{\odot}}$ and supermassive $M={10^8~{M}_{\odot}}$ black holes. Test particles with different specific charges $q/m$ ratios are assumed - from electron and proton to charged dust grain (one electron lost, $m=2\times10^{-18}$~kg).
\label{tabX}
} 
\end{center}
\end{table}

The range of the magnetic field magnitude for $\cb=0.01$ magnetic parameter 
and stellar $M={10~{M}_{\odot}}$, or supermassive $M={10^8~{M}_{\odot}}$, BH and various specific charges of ionized material from accretion disk is quite wide and hence it should not be complicated to find astrophysically relevant situation for presented model of ionized disk. Even weak magnetic field, like for example Galactic magnetic fields $10^{-5}~{\rm{G}}$, could have strong impact on ionized accretion disks. Another point to address is the difference of Lorentz force magnitude action on electron and proton. For the same magnitude of magnetic field $B$ in Gauss, the electron magnetic field parameter $\cb_{\rm e}$ will be 1836 time bigger then the proton magnetic field parameter $\cb_{\rm p}$, hence we can assume the protons should stay in accretion disk plane while electrons will escape disk plane along magnetic field lines in properly tuned magnetic field.


\section{Conclusions}

When the inclination of the magnetic field to the neutral Keplerian disk is almost perpendicular, the ionized particle motion can be almost completely regular. As the magnetic field inclination to the field decreases, the charged particle motion becomes chaotic with chaos-measure increasing with decreasing inclination to middle angles, until some limiting value of the inclination after which the charged particle is captured by the black hole.  
The most chaotic behaviour of the ionized particles, and the originally Keplerian disk destruction can be expected for magnetic field strength $|\cb|\in(0.01,1)$, weaker magnetic fields, with $\cb\in(0,0.01)$, are not strong enough to destroy the disk. The very strong magnetic fields, with $|\cb|\geq1$, which is from the astrophysics points of view also the most relevant case for the elementary particles, will take complete control over the particle dynamics due to the dominating Lorentz force, and the charged particles mainly follow the magnetic field lines.

All presented non-linear methods have been used for chaos determination in the charged particle motion - all methods are giving similar results, but the CPU time consumption differs for different methods. This could be due to algorithm programming, but it is dominantly caused by the fact that different numerical methods have different computational complexity requiring different computational times. One of the most promising methods is the Machine Learning, which can determine distinction between regular/chaotic trajectory very fast. 
All the considered methods are giving similar results while measuring the chaos in the charged particle motion -- the structures presented in Fig. \ref{figComparison} are similar. They have different scales for the chaos-measure and thus they also differ in the colour palette reflecting different ranges of the chaos measure given by different non-linear methods. 

Variety of phenomena related to the combined gravo-magnetic effects on charged test particle motion around rotating Kerr black holes is much richer than for the \Schw{} black holes, especially, the black hole rotation allows the ionized particles to escape to infinity along the direction of the external magnetic field \cite{Stu-Kol:2016:EPJC:}. The study of charged particles 'kicked' from the innermost stable circular orbit (ISCO) in the equatorial plane, and hence escaping to infinity, has been treated in \cite{Zah-etal:2013:PHYSR4:,Shi-Kim-Chi:2014:PHYSR4:,Zah:PRD:2014:}. Actually, for some large enough magnetic field parameter $|\cb|\geq1$, the charged particles must escape from the equatorial plane even with zero kick \cite{Stu-Kol:2016:EPJC:}. The reason for such an effect is that the particle orbit is stable in the radial $r/x$ direction, being unstable in the vertical $\theta/z$ direction, and the particle is forced to escape in the vertical direction -- see the second trajectory in Fig. 11 from \cite{Tur-Stu-Kol:2016:PHYSR4:}. The exploration of MPP effect in such configuration, and the possibility of the black hole rotational energy extraction due to the discharge of the Wald black hole charge is an interesting new phenomenon governing acceleration of the ultra-relativistic jets \cite{Tur-etal:2018:submitted}.

It will be also interesting to examine how the chaos of the charged particle motion will be modified, and how the ionized Keplerian disk will be influenced by the external magnetic field, when the charged particle radiation reaction \cite{Tur-etal:APJ:2018:} will be taken into account. As one can expect, the radiation reaction force will act as dumping in harmonic oscillator, forcing the chaotic particle trajectory to become more regular. As related to the observational data, we have to recall the argument that the signal from the accretion disks orbiting neutron stars with magnetic field having much larger magnitude than those related to black holes tends to be more chaotic in comparison with the signal from the accretion disks orbiting magnetized black holes \cite{Kar-Dut-Muk:APJ:2010:}. We plan to develop both detailed theoretical models of observational effects related to the ionized accretion disks and more complex structures, and further testing of non-linear methods of measuring the level of chaotic motion applied in the more detailed treatment of theoretical data, and the related real observational data. 

\section*{Acknowledgments}

The authors acknowledge the Silesian University in Opava Grant No. SGS/14/2016. M.K. acknowledges the Czech Science Foundation Grant No. 16-03564Y, Z.S. acknowledges the Albert Einstein Center for Gravitation and Astrophysics supported by the Czech Science Foundation Grant No. 14-37086G. 


\appendix


\begin{figure*}  
\includegraphics[width=\hsize]{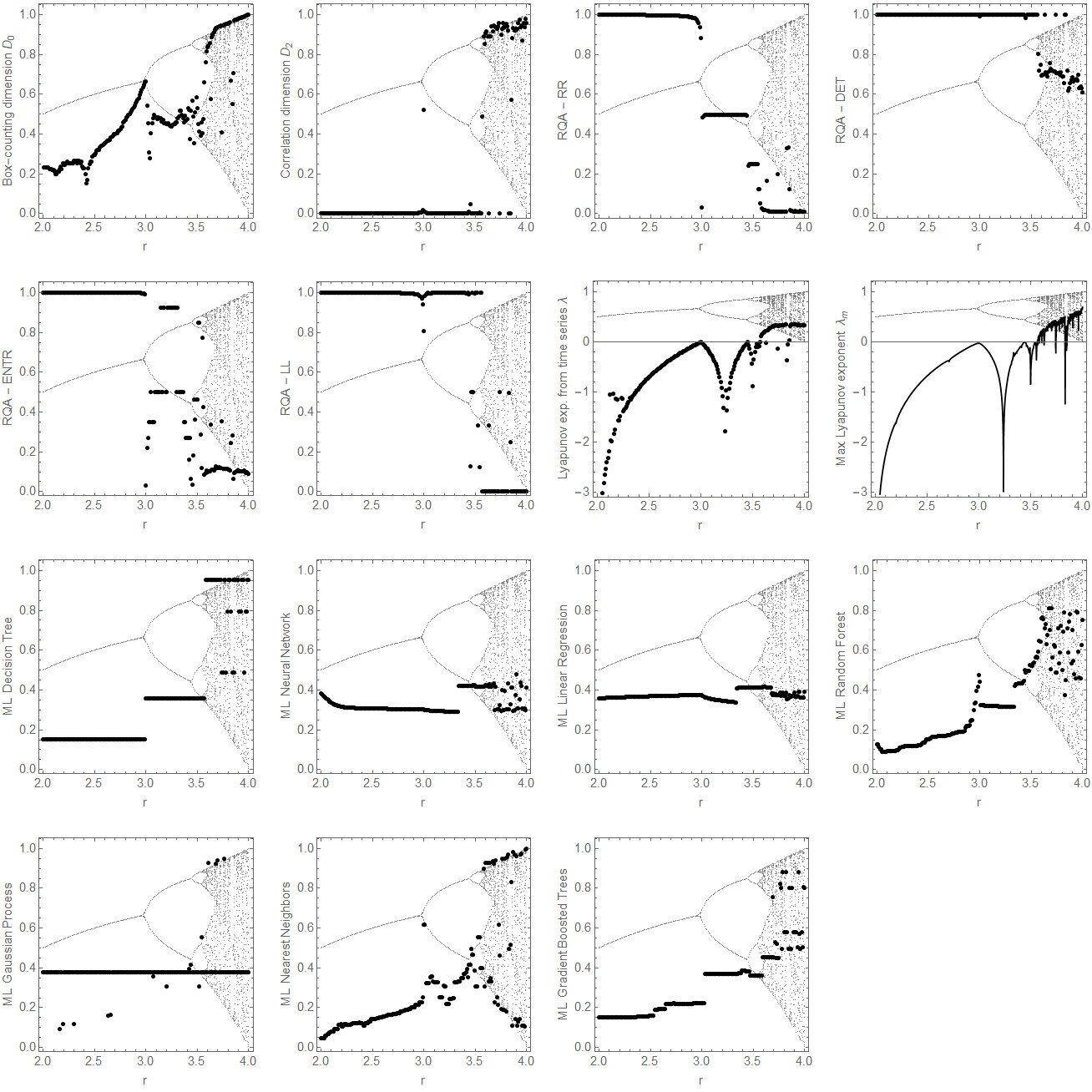}
\caption{\label{porovnanie}
Comparison of different nonlinear methods for time series generated by logistic map. For different parameter $r$ we generate 10000 points series with initial value use $x_0=0.1$.
}
\end{figure*}

\section{ Methods of determination of chaotic behaviour in time series \label{section_logistic}}


The motion of charged particles in the field of a magnetized black hole demonstrates a mixture of regularity and deterministic chaos. Deterministic chaos is a hardly predictable and apparently random behaviour which can appear in dynamical systems, being detectable by non-linear methods only. We first present a short survey of the methods detecting the deterministic chaos, and transitions between the regular and chaotic behaviour of dynamical systems, and then we test these methods in the case of so called logistic map representing one of the simplest dynamical systems. 

\subsection{Methods detecting chaos}
Dynamical systems, represented by time series of data, can demonstrate regular behaviour, completely random behaviour, or the so called deterministic chaos. The standard linear methods as the Fourier analysis are not relevant in distinguishing these three types of data, therefore, non-linear methods of treating such data samples are necessary. Such methods are able to give even a measure of chaos in the detected data. Here we give a survey of these methods. 

\subsubsection{Box-counting \label{boxcount}}

Box-counting ($D_0$) or box dimension is one of the most widely used methods of estimations of fractal dimension. The calculation and empirical application of this method is simple as compared to other methods. We present here the general idea behind this method, for detailed description see \cite{1991fcpl.book.....S}. For a set $ S $ in a Euclidean space $ \mathbb{R}^n  $, we define the Box-counting measure as
\beq
D_0 = \lim_{\epsilon \to 0}  \frac{\ln N(\epsilon)}{ \ln \frac{1}{\epsilon}},
\eeq
where $N(\epsilon)$ is the number of boxes of side length  $\epsilon$ required to cover the set. The dimension $D_0$ of the set $S$ is estimated by seeing how the logarithmic rate of $N(\epsilon)$ increases as ${\epsilon \to 0}$, or in other words, as we make the grid finer.

\subsubsection{Correlation dimension  \label{correlation}}

Very popular tool for detecting chaos in experimental data is calculation of the Correlation dimension ($D_2$). The general idea behind computing correlation dimension is to find out for some small $\epsilon $  the number of points $ C(\epsilon)$ (correlation sum),  which Euclidean distance is smaller than $\epsilon$. 
The definition of the Correlation dimension takes the form 
\beq
D_2 = \lim_{\epsilon \to 0}  \frac{\ln C(\epsilon)}{ \ln \epsilon},
\eeq
and one can compute the number $C$ for various values of $\epsilon$ and $ D_2 $ can be approximated again by fitting the logarithmic values. 

It is worth to mention that $ D_2 $  and  $ D_0 $  are subclasses of the general definition of $D_q $ determining the family of fractal dimensions \cite{1991fcpl.book.....S}. The fractal dimensions are defined by the relation  
\beq
D_q =    \lim_{\epsilon \to 0} \frac{1}{q - 1}  \frac{\ln \sum_k p ^q _k }{ \ln \epsilon}   \quad   - \infty \leq q \leq \infty,
\eeq
where, $p_k$ denotes relative frequency with which the fractal points are falling inside the k-th cell.

For $ q = 0 $ we obtain the Box-counting dimension, for $ q  \to 1 $ we obtain the so called information dimension, the numerator of which is denoted as Shannon's entropy, and for  $ q = 2 $ we obtain the Correlation dimension.

There is several approaches to the calculation procedure of the correlation dimension -- let us mention the approximation of $C(\epsilon) $ by the algorithm of Grassberger and Procaccia \cite{1983PhRvL..50..346G}
\beq
\hat{C}(\epsilon)= \lim_{N \to \infty} \frac{2}{N ( N - 1 )} \mathop{\sum_{i < j  }} H ( \epsilon - | x_i - x_j |),
\eeq
where $H$ is the Heaviside step function. 
 
When counting correlation dimension (fractal dimension), one should not omit importance of the embedding dimension, and also the close connection to Takens's theorem about reconstruction of state space from sequence of observations. The embedding dimension is created from series of length $N + m - 1 $ for some given $m$ series of $N$ vectors, where i-th component takes the form
\beq
x_i=(x_{i-m+1}, x_{i-m+2}, ...,x_i) \in \mathbb{R}^m .
\eeq
One of the purposes of the embedding dimension is to distinguish between chaotic and random time series. In the chaotic series, for increasing $m \in \mathbb{N}$, the fractal dimension stabilizes at some value $D < m$, while in the random series, the fractal dimension goes along with $m$ to infinity. 

\subsubsection{Lyapunov exponent \label{lyapunov}}


In short, a Lyapunov exponent is a number giving the measure of separation of trajectories that are initially infinitesimally close. Near trajectories in chaotic systems diverge exponentially, what corresponds to positive Lyapunov exponents. The amount of separation can differ for various directions of the initial separation vector. Because of this fact, there is a spectrum of the Lyapunov exponents, which corresponds to the phase space dimension.

For example, if we consider the logistic map $ f(x) = r x(1-x), r \in [0,4]$ as a typical example of simple chaotic system, which represents one-dimensional non-linear difference equation, the maximal Lyapunov exponent can be calculated directly from the expression of the $f$ function from the formula 
\beq \label{lyap}
\lambda_m = \lim_{N \to \infty} \frac{1}{N}  \mathop{\sum_{n = 0 }}_{}^{N-1} \ln [ \, | f'(f^n (x_0)) | \,  ].  
\eeq
However, we assume here handling with time series inputs only, forbidding application of such a formula. This approach leads us to use method, which is determining the Lyapunov exponents from time series, being described in \cite{1994PhLA..185...77K}. The maximal Lyapunov exponent characterizes the spectrum, and therefore denotes amount of predictability for the considered dynamical system \footnote{Note that the Lyapunov coefficients have a crucial role also in other branches of physics, e.g., they characterize instability of quasinormal modes of perturbation fields around black holes \citep{2009PhRvD..79f4016C,2017PhLB..771..597K, 2018PhRvD..97h4058T}.}. 

The Lyapunov exponents can be calculated without knowledge of a model behind the dynamical system, being based only on the observed time series. We use the method based on the statistical properties of the divergence of the neighbouring trajectories, introduced by Kantz (1994) \cite{1994PhLA..185...77K}. We denote our numerical estimation by $\lambda$, which is actually sophisticated way of estimating true Lyapunov exponent as the average of maximal effective exponents  $\lambda_\tau$ along the trajectory.  The algorithm is well explained in Kantz's article,  it is quite fast and powerful method. The result of this numerical approach   applied to logistic map is very similar to the result of \ref{lyap} as one can compare in Fig. \ref{porovnanie}. The observable differences appears in this case  by non-chaotic or even constant time series, which are negligible for the purposes of our study. 


\subsubsection{Recurrence quantification analysis \label{xRQAx}}

The recurrence quantification analysis (RQA) is a widely used tool for investigation of the state space trajectories -- it determines the number and duration of recurrences of a dynamical system. RQA was developed since 1992 by Zbilut, Webber Jr. and Marwan \cite{1992PhLA..171..199Z,2008EPJST.164....3M}. Recurrence plots (RP) provide a graphical tool for observing periodicity of phase space trajectories and was introduced by Eckmann et al.in 1987 \cite{1987EL......4..973E}. This observing is possible through visualization of a symmetrical square matrix, in which the elements correspond to times at which a state of a dynamical system recurs.

We can define a RP which measures recurrences of a trajectory $ x_i \in  R^d$  in phase space
\beq
R_{i,j}= H(\epsilon - \| x_i - x_j\|  \mid ) \quad  i, j = 1, ...,N,
\eeq
where $N$ is the number of measured points $x_i$, $\epsilon$ is a threshold distance, and $ \| \cdot \|$ is a norm.  From this equation we obtain the already mentioned symmetrical square matrix of zeroes and ones. When we represent the two repeating elements with different colours in a plot, we obtain the discussed RP. Threshold value parameter determines density of the RP plot.

For investigating of the chaotic trajectories we use the following RQA variant tools:
\begin{enumerate}
\item RR - The recurrence rate is simplest tool, measuring the density of the recurrence points in the RP -- in another words, it divides the number of the recurrence points in the RP by the number of all elements in the matrix. RR tool reflects the chance that some state of the system will recur
 \beq
 RR = \frac{1}{N^2} \sum_{i,j=1}^N R_{i,j}. 
 \eeq
\item  DET - Determinism is the rate of the recurrence points which build the diagonal lines. DET determines how predictable the system is being formally defined as 
\beq
DET  = \frac{\sum_{l=l_{min}}^N l P(l)}{ \sum_{i,j=1}^N R_{i,j}}, 
\eeq
where  $P(l)$ denotes the frequency distribution of the lengths $l$ of the diagonal lines.
\item  LL - Line length tool represents average diagonal line length, which is in relation with the time of predictability of the system. It reflects the average time for which any two parts of the considered trajectories are close -- this time can be denoted as mean prediction time 
\beq
LL = \frac{\sum_{l=l_{min}}^N l P(l)}{\sum_{l=l_{min}}^N  P(l)}.
\eeq
\item  ENTR - Entropy (or the Shannon entropy) of the probability distribution of the diagonal line lengths p(l), which are reflecting complexity of the system deterministic structure. The definition reads 
\beq
   ENTR =   - \sum_{l=l_{min}}^N  p(l) \ln p(l),
\eeq
where $p(l)$ is the probability that a diagonal line is exactly of the length $l$ -- it can be estimated from the frequency distribution $P(l)$ with \\
 $p(l) = \frac{P(l)}{\sum_{l=l_{min}}^N  P(l)}$. 
\end{enumerate}
 
\subsubsection{Machine learning  \label{machinelearning}}
Machine learning is a very powerful tool, which finds application in many quite diverse fields -- language translating algorithms, computer vision, beating best Go player in the world  \cite{2017Natur.550..354S}, or Chess programs with different architecture \cite{2017arXiv171201815S}. The purpose of this paper is to briefly introduce its basic principles and application to the study the chaotic motion. 

Roughly speaking, machine learning (ML) is a field of computer science, which is strongly connected with another fields of mathematics as optimization, statistics, linear algebra, etc. Its beginning goes to 1950's and as many inventions in computer science, or in science generally, the ML was not invented by a single person, but we mention at least A. Samuel, who used for the first time the term "Machine learning". The ML is using algorithms on data samples to discover known or unknown patterns in data. This dividing of patterns leads to basic divisions of the ML, namely the supervised, semi-supervised and unsupervised learning. The wide range of applications announces the good ability of handling non-linear dynamical systems data.

Our intention of using the ML method is to decide whether a trajectory of a charged particle is chaotic or not. For this purpose we use the supervised ML, where we train various ML algorithms with examples from the non-linear methods we have already described. The training set consists overall of 4725 examples. Chaos is very complex phenomenon which is reflected by various another phenomena like sensitivity to initial conditions, fractal dimension of coordinates time series, attractors, or the predictability of the dynamical system. We include into our training data set the results from all the already mentioned methods (Box-counting, Correlation dimension, Lyapunov exponent, RQA - RR, DET, LL, ENTR ) with effort to use all the different properties of them. The appropriate number of trajectories has been assigned to evenly weighted sum of previously calculated chaos-measure estimations based on previously mentioned methods. Various types of the ML algorithms has been trained with these data, the most of them are presented in Fig. \ref{figComparison}. 

The ML algorithms considered in our study are the following; the description is  exactly taken from the cited sources:
\begin{enumerate}
\item  Decision Tree \cite{M1} -- A decision tree is a flow chart–like structure in which each internal node represents a "test" on a feature, each branch represents the outcome of the test, and each leaf represents a class or value distribution.

\item  Neural Network \cite{M2} --  A neural network consists of stacked layers, each performing a simple computation. Information is processed layer by layer from the input layer to the output layer. The neural network is trained to minimize a loss function on the training set using gradient descent.

\item Linear Regression \cite{M3} -- The linear regression predicts the numerical output $y$ using a linear combination of numerical features. The conditional probability $P(y|x)$  is modeled according to
\beq P(y|x) \propto  exp(-(y - f(\theta,x))^2 / (2 \sigma)^2), \eeq with $ f(\theta,x) = x \cdot \theta $. 

The estimation of the parameter vector $\theta $ is done by minimizing the loss function 
\beq \frac{1}{2} \sum_{i=1}^{m}  (y_i  - f(\theta,x_i))^2 + \lambda_1 \sum_{i=1}^{n} |\theta_i | +  \frac{\lambda_2}{2}|\theta_i ^2,   \eeq
 where $m$ is the number of examples and $n$ is the number of numerical features.
 
\item  Random Forest \cite{M4} -- Random forest is an ensemble learning method for classification and regression that operates by constructing a multitude of decision trees. The forest prediction is obtained by taking the most common class or the mean-value tree predictions. Each decision tree is trained on a random subset of the training set and only uses a random subset of the features (bootstrap aggregating algorithm).

\item   Gaussian Process  \cite{M5} -- The "GaussianProcess" method assumes that the function to be modeled has been generated from a Gaussian process. The Gaussian process is defined by its covariance function (also called kernel). In the training phase, the method will estimate the parameters of this covariance function. The Gaussian process is then conditioned on the training data and used to infer the value of a new example using a Bayesian inference.

\item   Nearest Neighbours \cite{M6} --  Nearest neighbors is a type of instance-based learning. In its simplest form, it picks the commonest class or averages the values among the k nearest neighbors.

\item    Gradient Boosted Trees \cite{M7} --  Gradient boosting is a machine learning technique for regression and classification problems that produces a prediction model in the form of an ensemble of trees. Trees are trained sequentially with the goal of compensating the weaknesses of previous trees. The current implementation uses the LightGBM framework in the back end.
\end{enumerate} 

In the final stages of our study, represented by Fig. \ref{figMag}, we decided to compare the results of the Correlation dimension method with the efficient ML - Random Forest method, applied to the sample of $10^4$ pairs of initial conditions for various magnetic field parameters values. 


\subsection{Test of the non-linear methods detecting chaos}
Classical example of a simple non-linear dynamical system is the so called logistic map, given by the quadratic recurrence equation
\beq
 x_{n+1} = r x_{n}(1-x_{n}). \label{logistic}
\eeq
If the initial value of the variable is given, $x_0\in(0,1)$, the logistic map (\ref{logistic}) generates sequences of real numbers $x_n\in(0,1)$ in dependence on the map parameter $r$. The behaviour of such sequence $x_n$ strongly depends on logistic map parameter that is considered in the interval $r\in[2,4]$. Roughly speaking, the behaviour on the interval $r\in(2,r_0)$ is regular (rather predictable) and on the interval $r\in[r_0,4]$ it is chaotic (hardly predictable or rather unpredictable), with some occasional "islands of regularity". The transition between regular and chaotic behaviour happens for logistic map parameter $r=r_0\approx~3.56995$. Bifurcation diagram for the logistic map, with asymptotically approached values of the sequence, is shown in Fig. \ref{porovnanie} by gray background points.


For chaos determination and chaotic behaviour description in sequences of numbers, the following methods have been tested: Box-Counting (section \ref{boxcount}), Correlation Dimension (\ref{correlation}), Lyapunov Exponent (\ref{lyapunov}), Recurrence Quantification Analysis (RQA) with variants (\ref{xRQAx}), and Machine Learning with various algorithms (\ref{machinelearning}). All of the algorithms we use in this article are capable to be applied for one dimensional sequences of real numbers (time series) taken as an input. From the point of observation, it is important to distinguish between chaotic sequences ruled by some (unknown) laws, and random sequences, where the numbers are chosen randomly. The theoretical boundary of distinguishing between chaos and random sequences is sequence length \cite{Ott:1993:book:}. For the non-linear dynamical system having many degrees of freedom, short sequences are not sufficient for any decision.


To clear up how the non-linear methods work, we use four representative sequences of length $10^4$
\begin{equation}
\begin{split}
 x_n^{(1)}  & = \{ 0.842154, 0.451963, 0.842154, 0.451963, \ldots \} \\ 
 x_n^{(2)}  & = \{  0.1, 0.324, 0.788486, 0.600392, 0.863717, \ldots \} \\
 x_n^{(3)}   & = \{ 0.1, 0.36, 0.9216, 0.289014, 0.821939,  \ldots \} \\
 x_n^{(4)}   &= \{ 0.180463, 0.757133, 0.699205, 0.535334,   \ldots \}, 
 \end{split}
\end{equation}
where $ x_n^{(1)}$ is a regular sequence, $ x_n^{(2)}$ is a weakly chaotic sequence generated by the logistic map with $r = 3.6 $, $x_n^{(3)}$ is a strongly chaotic sequence generated by the logistic map with $ r = 4 $, and $x_n^{(4)}$ is a sequence of pseudo-random distribution of numbers. The results are presented in Tab. \ref{tab1}. 

We have tested in detail all the introduced non-linear methods of chaos determination on sequences of numbers generated by the logistic map for various values of the parameter $r$ -- for results see Fig. \ref{porovnanie}. The important common sign of all these methods is that they are able to detect higher chaos-measure when divergence of the trajectories in the bifurcation point of the diagram at $r=3$ occurs, i.e., when there is more than one fixed point. However, this is not undeniable true for all the cases, what leads to different estimations of the chaos-measure for different non-linear methods. The Box-counting and Lyapunov exponent methods show approximately linear behaviour when moving to the point, where the divergence of the trajectories begins -- while moving to this point, the chaos-measure is increasing, and it is decreasing while leaving it. The Correlation dimension method demonstrates little bit different behaviour, where non-chaotic region is denoted by values close to zero, and the level of chaos starts to rapidly grow only in small distance from the region of divergence of trajectories in the bifurcation point -- this type of the behaviour is also demonstrated by the RQA tools, however, in a reversed way. The determination of chaotic behaviour for $r \geq 3.56995 $, when all the non-linear methods are showing the highest values of the chaos-measure, can be considered as highly satisfactory. However, as in this region also the "islands of regularity" are present, not all of the chaos-measure estimations should be high there.

As expected, different ML algorithms produce different results. The Random Forest algorithm, used also in \citep{Hup-etal:MNRAS:2017:} seems to produce quite reliable results -- our estimation of the chaos-measure is getting higher as the $r$ parameter grows, with exceptions that could be explained by the fact there are also "islands of regularity" in the chaotic region. This is the reason why we decided to include the ML Random Forest algorithm to the figure \ref{figMag}; but we have to stress that the others ML algorithms seem to produce interesting and relevant results as well.  

The time spent for the calculations of every single figure from Fig. \ref{porovnanie} is presented in Tab. \ref{tab2}. From here we can see that the Box-counting method is the fastest, but when looking on the results, the other non-linear methods seem to be more precise. The time spent on the calculations also strongly depends on the sequence length - such dependence is non-linear in the case of almost all the methods, but we should point out that Box-counting and ML seem to behave generally in rather linear way while working with larger data sets (however, this general conclusion is not fully confirmed by the results presented in Tab \ref{tab2}). 

For most of the non-linear chaos detecting methods we have programmed and tested several variants according to the algorithm architecture, and settings of the adjustable parameters, for the code see \cite{Git}. The goal was to obtain relevant results and acceptable computing time.

\begin{table}
\caption{\label{tab1} 
Comparison of values of different nonlinear methods applied on regular, chaotic and pseudo-random generated series of the length 10 000. }
\begin{ruledtabular}
\begin{center}
    \begin{tabular}{| l | l | l | l | l |}
    \hline
    Met./ Type & Regular & Weak ch. & Strong ch. & P.-rand.\\ \hline
   Box-count & 0.0969 & 0.8343 & 1   & 1 \\ \hline
   Corr. dim. & $5*10^{-15}$ & 0.8943 & 0.8735 & 0.9978  \\ \hline
    Lyap. exp. &  $-7*10^{-16}$ & 0.1889 & 0.3289  &   0.0842  \\    \hline
  RQA-RR & 0.4999 & 0.0183 & 0.0116 & 0.0057 \\    \hline
   RQA-DET & 1 & 0.6959 & 0.6114   &  0.0113 \\    \hline
   RQA-ENTR & 8.5152  & 0.8951 & 0.8202   &   0.0296 \\    \hline
   RQA-LAM & 5001 & 5.6299 & 2.8507  &  2.0006 \\    \hline
     ML-Rand. For. & 0.5915 & 0.6917 & 0.7869  & 0.3788  \\    \hline
    \end{tabular}
\end{center}
\end{ruledtabular}
\end{table}
\begin{table}
\caption{\label{tab2} Comparison of 
time in seconds required for different nonlinear methods applied on  time series generated by logistic map. For  parameter $r$ varying from $ 2$ to $4$ with the step of $0.01$ leads to $200$ time series, with initial value  $x_0=0.1$ we did set up the iterations to length  $ 100, 1000$ and $ 10000 $ and compared the time needed for the calculation  for given methods.}
\begin{ruledtabular}
\begin{center}
    \begin{tabular}{| l | l | l | l |}
    \hline
    Method / Length & 100 & 1000 & 10000\\ \hline
   Box-count & 0.354 & 0.416 & 3.519  \\ \hline
   Correlation dim. & 4.155& 26.54 & 2353 \\ \hline
    Lyapunov exp. & 3.294 & 119.5 & 12574 \\    \hline
  RQA- RR & 0.164 & 6.332 & 507.8 \\    \hline
   RQA-DET & 0.209 & 6.069 &  916.9 \\    \hline
   RQA-ENTR &  0.197 &  7.247 &  1128  \\    \hline
   RQA-LAM & 0.197 & 7.248 & 1128 \\    \hline
       ML-Rand. For. & 0.677 & 1.212 & 7.479 \\    \hline
    \end{tabular}
\end{center}
\end{ruledtabular}
\end{table}



%
%


\begin{thebibliography}{60}
\expandafter\ifx\csname natexlab\endcsname\relax\def\natexlab#1{#1}\fi
\expandafter\ifx\csname bibnamefont\endcsname\relax
  \def\bibnamefont#1{#1}\fi
\expandafter\ifx\csname bibfnamefont\endcsname\relax
  \def\bibfnamefont#1{#1}\fi
\expandafter\ifx\csname citenamefont\endcsname\relax
  \def\citenamefont#1{#1}\fi
\expandafter\ifx\csname url\endcsname\relax
  \def\url#1{\texttt{#1}}\fi
\expandafter\ifx\csname urlprefix\endcsname\relax\def\urlprefix{URL }\fi
\providecommand{\bibinfo}[2]{#2}
\providecommand{\eprint}[2][]{\url{#2}}

\bibitem[{\citenamefont{{Carter}}(1968)}]{Car:1968:CMaPh:}
\bibinfo{author}{\bibfnamefont{B.}~\bibnamefont{{Carter}}},
  \bibinfo{journal}{Communications in Mathematical Physics}
  \textbf{\bibinfo{volume}{10}}, \bibinfo{pages}{280} (\bibinfo{year}{1968}).

\bibitem[{\citenamefont{{Misner} et~al.}(1973)\citenamefont{{Misner}, {Thorne},
  and {Wheeler}}}]{Mis-Tho-Whe:1973:Gravitation:}
\bibinfo{author}{\bibfnamefont{C.~W.} \bibnamefont{{Misner}}},
  \bibinfo{author}{\bibfnamefont{K.~S.} \bibnamefont{{Thorne}}},
  \bibnamefont{and} \bibinfo{author}{\bibfnamefont{J.~A.}
  \bibnamefont{{Wheeler}}}, \emph{\bibinfo{title}{{Gravitation}}}
  (\bibinfo{year}{1973}).

\bibitem[{\citenamefont{{Bi{\v{c}}{\'a}k}
  et~al.}(1989)\citenamefont{{Bi{\v{c}}{\'a}k}, {Stuchl{\'{\i}}k}, and
  {Balek}}}]{Bic-Stu-Bal:1989:BAC:}
\bibinfo{author}{\bibfnamefont{J.}~\bibnamefont{{Bi{\v{c}}{\'a}k}}},
  \bibinfo{author}{\bibfnamefont{Z.}~\bibnamefont{{Stuchl{\'{\i}}k}}},
  \bibnamefont{and} \bibinfo{author}{\bibfnamefont{V.}~\bibnamefont{{Balek}}},
  \bibinfo{journal}{Bulletin of the Astronomical Institutes of Czechoslovakia}
  \textbf{\bibinfo{volume}{40}}, \bibinfo{pages}{65} (\bibinfo{year}{1989}).

\bibitem[{\citenamefont{{Blaschke} and
  {Stuchl{\'{\i}}k}}(2016)}]{Bla-Stu:2016:PYSR4:..94h6006B}
\bibinfo{author}{\bibfnamefont{M.}~\bibnamefont{{Blaschke}}} \bibnamefont{and}
  \bibinfo{author}{\bibfnamefont{Z.}~\bibnamefont{{Stuchl{\'{\i}}k}}},
  \bibinfo{journal}{\prd} \textbf{\bibinfo{volume}{94}}, \bibinfo{eid}{086006}
  (\bibinfo{year}{2016}), \eprint{1610.07462}.

\bibitem[{\citenamefont{{Carter}}(1973)}]{Car:1973:BlaHol:}
\bibinfo{author}{\bibfnamefont{B.}~\bibnamefont{{Carter}}}, in
  \emph{\bibinfo{booktitle}{Black Holes (Les Astres Occlus)}}, edited by
  \bibinfo{editor}{\bibfnamefont{C.}~\bibnamefont{{Dewitt}}} \bibnamefont{and}
  \bibinfo{editor}{\bibfnamefont{B.~S.} \bibnamefont{{Dewitt}}}
  (\bibinfo{year}{1973}), pp. \bibinfo{pages}{57--214}.

\bibitem[{\citenamefont{Stuchl{\'{\i}}k}(1983)}]{Stu:1983:BULAI:}
\bibinfo{author}{\bibfnamefont{Z.}~\bibnamefont{Stuchl{\'{\i}}k}},
  \bibinfo{journal}{Bulletin of the Astronomical Institutes of Czechoslovakia}
  \textbf{\bibinfo{volume}{34}}, \bibinfo{pages}{129} (\bibinfo{year}{1983}).

\bibitem[{\citenamefont{{Karas} and
  {Vokrouhlick{\'y}}}(1992)}]{1992GReGr..24..729K}
\bibinfo{author}{\bibfnamefont{V.}~\bibnamefont{{Karas}}} \bibnamefont{and}
  \bibinfo{author}{\bibfnamefont{D.}~\bibnamefont{{Vokrouhlick{\'y}}}},
  \bibinfo{journal}{General Relativity and Gravitation}
  \textbf{\bibinfo{volume}{24}}, \bibinfo{pages}{729} (\bibinfo{year}{1992}).

\bibitem[{\citenamefont{{Stuchl{\'{\i}}k} and {Kolo{\v
  s}}}(2016)}]{Stu-Kol:2016:EPJC:}
\bibinfo{author}{\bibfnamefont{Z.}~\bibnamefont{{Stuchl{\'{\i}}k}}}
  \bibnamefont{and} \bibinfo{author}{\bibfnamefont{M.}~\bibnamefont{{Kolo{\v
  s}}}}, \bibinfo{journal}{European Physical Journal C}
  \textbf{\bibinfo{volume}{76}}, \bibinfo{eid}{32} (\bibinfo{year}{2016}),
  \eprint{1511.02936}.

\bibitem[{\citenamefont{{Frolov} and {Shoom}}(2010)}]{Fro-Sho:2010:PHYSR4:}
\bibinfo{author}{\bibfnamefont{V.~P.} \bibnamefont{{Frolov}}} \bibnamefont{and}
  \bibinfo{author}{\bibfnamefont{A.~A.} \bibnamefont{{Shoom}}},
  \bibinfo{journal}{\prd} \textbf{\bibinfo{volume}{82}}, \bibinfo{eid}{084034}
  (\bibinfo{year}{2010}), \eprint{1008.2985}.

\bibitem[{\citenamefont{{Kov{\'a}{\v r}}
  et~al.}(2008)\citenamefont{{Kov{\'a}{\v r}}, {Stuchl{\'{\i}}k}, and
  {Karas}}}]{Kov-Stu-Kar:2008:CLAQG:}
\bibinfo{author}{\bibfnamefont{J.}~\bibnamefont{{Kov{\'a}{\v r}}}},
  \bibinfo{author}{\bibfnamefont{Z.}~\bibnamefont{{Stuchl{\'{\i}}k}}},
  \bibnamefont{and} \bibinfo{author}{\bibfnamefont{V.}~\bibnamefont{{Karas}}},
  \bibinfo{journal}{Classical and Quantum Gravity}
  \textbf{\bibinfo{volume}{25}}, \bibinfo{eid}{095011} (\bibinfo{year}{2008}),
  \eprint{0803.3155}.

\bibitem[{\citenamefont{{Kolo{\v s}} et~al.}(2017)\citenamefont{{Kolo{\v s}},
  {Tursunov}, and {Stuchl{\'{\i}}k}}}]{Kol-Tur-Stu:2017:EPJC:}
\bibinfo{author}{\bibfnamefont{M.}~\bibnamefont{{Kolo{\v s}}}},
  \bibinfo{author}{\bibfnamefont{A.}~\bibnamefont{{Tursunov}}},
  \bibnamefont{and}
  \bibinfo{author}{\bibfnamefont{Z.}~\bibnamefont{{Stuchl{\'{\i}}k}}},
  \bibinfo{journal}{European Physical Journal C} \textbf{\bibinfo{volume}{77}},
  \bibinfo{eid}{860} (\bibinfo{year}{2017}), \eprint{1707.02224}.

\bibitem[{\citenamefont{{Wald}}(1974)}]{Wal:1974:PHYSR4:}
\bibinfo{author}{\bibfnamefont{R.~M.} \bibnamefont{{Wald}}},
  \bibinfo{journal}{\prd} \textbf{\bibinfo{volume}{10}}, \bibinfo{pages}{1680}
  (\bibinfo{year}{1974}).

\bibitem[{\citenamefont{{Prasanna} and
  {Vishveshwara}}(1978)}]{Pra-Vis:1978:Pra:}
\bibinfo{author}{\bibfnamefont{A.~R.} \bibnamefont{{Prasanna}}}
  \bibnamefont{and}
  \bibinfo{author}{\bibfnamefont{V.}~\bibnamefont{{Vishveshwara}}},
  \bibinfo{journal}{Pramana} \textbf{\bibinfo{volume}{11}},
  \bibinfo{pages}{359} (\bibinfo{year}{1978}).

\bibitem[{\citenamefont{{Prasanna}}(1980)}]{Prasanna:1980:RDNC:}
\bibinfo{author}{\bibfnamefont{A.~R.} \bibnamefont{{Prasanna}}},
  \bibinfo{journal}{Nuovo Cimento Rivista Serie} \textbf{\bibinfo{volume}{3}},
  \bibinfo{pages}{1} (\bibinfo{year}{1980}).

\bibitem[{\citenamefont{{Aliev} and {Galtsov}}(1981)}]{Ali-Gal:1981:GRG:}
\bibinfo{author}{\bibfnamefont{A.~N.} \bibnamefont{{Aliev}}} \bibnamefont{and}
  \bibinfo{author}{\bibfnamefont{D.~V.} \bibnamefont{{Galtsov}}},
  \bibinfo{journal}{General Relativity and Gravitation}
  \textbf{\bibinfo{volume}{13}}, \bibinfo{pages}{899} (\bibinfo{year}{1981}).

\bibitem[{\citenamefont{{Kov{\'a}{\v r}}
  et~al.}(2010)\citenamefont{{Kov{\'a}{\v r}}, {Kop{\'a}{\v c}ek}, {Karas}, and
  {Stuchl{\'{\i}}k}}}]{Kov-Kop-Kar-Stu:2010:CLAQG:}
\bibinfo{author}{\bibfnamefont{J.}~\bibnamefont{{Kov{\'a}{\v r}}}},
  \bibinfo{author}{\bibfnamefont{O.}~\bibnamefont{{Kop{\'a}{\v c}ek}}},
  \bibinfo{author}{\bibfnamefont{V.}~\bibnamefont{{Karas}}}, \bibnamefont{and}
  \bibinfo{author}{\bibfnamefont{Z.}~\bibnamefont{{Stuchl{\'{\i}}k}}},
  \bibinfo{journal}{Classical and Quantum Gravity}
  \textbf{\bibinfo{volume}{27}}, \bibinfo{eid}{135006} (\bibinfo{year}{2010}),
  \eprint{1005.3270}.

\bibitem[{\citenamefont{{Kop{\'a}{\v c}ek}
  et~al.}(2010)\citenamefont{{Kop{\'a}{\v c}ek}, {Karas}, {Kov{\'a}{\v r}}, and
  {Stuchl{\'{\i}}k}}}]{Kop-etal:2010:APJ:}
\bibinfo{author}{\bibfnamefont{O.}~\bibnamefont{{Kop{\'a}{\v c}ek}}},
  \bibinfo{author}{\bibfnamefont{V.}~\bibnamefont{{Karas}}},
  \bibinfo{author}{\bibfnamefont{J.}~\bibnamefont{{Kov{\'a}{\v r}}}},
  \bibnamefont{and}
  \bibinfo{author}{\bibfnamefont{Z.}~\bibnamefont{{Stuchl{\'{\i}}k}}},
  \bibinfo{journal}{\apj} \textbf{\bibinfo{volume}{722}}, \bibinfo{pages}{1240}
  (\bibinfo{year}{2010}), \eprint{1008.4650}.

\bibitem[{\citenamefont{{Abdujabbarov}
  et~al.}(2013{\natexlab{a}})\citenamefont{{Abdujabbarov}, {Ahmedov}, and
  {Jurayeva}}}]{Abd-etal:2013:PHYSR4:}
\bibinfo{author}{\bibfnamefont{A.~A.} \bibnamefont{{Abdujabbarov}}},
  \bibinfo{author}{\bibfnamefont{B.~J.} \bibnamefont{{Ahmedov}}},
  \bibnamefont{and} \bibinfo{author}{\bibfnamefont{N.~B.}
  \bibnamefont{{Jurayeva}}}, \bibinfo{journal}{\prd}
  \textbf{\bibinfo{volume}{87}}, \bibinfo{eid}{064042}
  (\bibinfo{year}{2013}{\natexlab{a}}).

\bibitem[{\citenamefont{{Al Zahrani} et~al.}(2013)\citenamefont{{Al Zahrani},
  {Frolov}, and {Shoom}}}]{Zah-etal:2013:PHYSR4:}
\bibinfo{author}{\bibfnamefont{A.~M.} \bibnamefont{{Al Zahrani}}},
  \bibinfo{author}{\bibfnamefont{V.~P.} \bibnamefont{{Frolov}}},
  \bibnamefont{and} \bibinfo{author}{\bibfnamefont{A.~A.}
  \bibnamefont{{Shoom}}}, \bibinfo{journal}{\prd}
  \textbf{\bibinfo{volume}{87}}, \bibinfo{eid}{084043} (\bibinfo{year}{2013}),
  \eprint{1301.4633}.

\bibitem[{\citenamefont{{Abdujabbarov}
  et~al.}(2013{\natexlab{b}})\citenamefont{{Abdujabbarov}, {Ahmedov}, and
  {Jurayeva}}}]{2013PhRvD..87f4042A}
\bibinfo{author}{\bibfnamefont{A.~A.} \bibnamefont{{Abdujabbarov}}},
  \bibinfo{author}{\bibfnamefont{B.~J.} \bibnamefont{{Ahmedov}}},
  \bibnamefont{and} \bibinfo{author}{\bibfnamefont{N.~B.}
  \bibnamefont{{Jurayeva}}}, \bibinfo{journal}{\prd}
  \textbf{\bibinfo{volume}{87}}, \bibinfo{eid}{064042}
  (\bibinfo{year}{2013}{\natexlab{b}}).

\bibitem[{\citenamefont{{Kop{\'a}{\v c}ek} and
  {Karas}}(2014)}]{Kop-Kar:2014:APJ:}
\bibinfo{author}{\bibfnamefont{O.}~\bibnamefont{{Kop{\'a}{\v c}ek}}}
  \bibnamefont{and} \bibinfo{author}{\bibfnamefont{V.}~\bibnamefont{{Karas}}},
  \bibinfo{journal}{\apj} \textbf{\bibinfo{volume}{787}}, \bibinfo{eid}{117}
  (\bibinfo{year}{2014}), \eprint{1404.5495}.

\bibitem[{\citenamefont{{Shiose} et~al.}(2014)\citenamefont{{Shiose}, {Kimura},
  and {Chiba}}}]{Shi-Kim-Chi:2014:PHYSR4:}
\bibinfo{author}{\bibfnamefont{R.}~\bibnamefont{{Shiose}}},
  \bibinfo{author}{\bibfnamefont{M.}~\bibnamefont{{Kimura}}}, \bibnamefont{and}
  \bibinfo{author}{\bibfnamefont{T.}~\bibnamefont{{Chiba}}},
  \bibinfo{journal}{\prd} \textbf{\bibinfo{volume}{90}}, \bibinfo{eid}{124016}
  (\bibinfo{year}{2014}), \eprint{1409.3310}.

\bibitem[{\citenamefont{{Kolo{\v s}} et~al.}(2015)\citenamefont{{Kolo{\v s}},
  {Stuchl{\'{\i}}k}, and {Tursunov}}}]{Kol-Stu-Tur:2015:CLAQG:}
\bibinfo{author}{\bibfnamefont{M.}~\bibnamefont{{Kolo{\v s}}}},
  \bibinfo{author}{\bibfnamefont{Z.}~\bibnamefont{{Stuchl{\'{\i}}k}}},
  \bibnamefont{and}
  \bibinfo{author}{\bibfnamefont{A.}~\bibnamefont{{Tursunov}}},
  \bibinfo{journal}{Classical and Quantum Gravity}
  \textbf{\bibinfo{volume}{32}}, \bibinfo{eid}{165009} (\bibinfo{year}{2015}),
  \eprint{1506.06799}.

\bibitem[{\citenamefont{{Tursunov} et~al.}(2016)\citenamefont{{Tursunov},
  {Stuchl{\'{\i}}k}, and {Kolo{\v s}}}}]{Tur-Stu-Kol:2016:PHYSR4:}
\bibinfo{author}{\bibfnamefont{A.}~\bibnamefont{{Tursunov}}},
  \bibinfo{author}{\bibfnamefont{Z.}~\bibnamefont{{Stuchl{\'{\i}}k}}},
  \bibnamefont{and} \bibinfo{author}{\bibfnamefont{M.}~\bibnamefont{{Kolo{\v
  s}}}}, \bibinfo{journal}{\prd} \textbf{\bibinfo{volume}{93}},
  \bibinfo{eid}{084012} (\bibinfo{year}{2016}), \eprint{1603.07264}.

\bibitem[{\citenamefont{{Kop{\'a}{\v c}ek} and
  {Karas}}(2018)}]{Kop-Kar:APJ:2018:}
\bibinfo{author}{\bibfnamefont{O.}~\bibnamefont{{Kop{\'a}{\v c}ek}}}
  \bibnamefont{and} \bibinfo{author}{\bibfnamefont{V.}~\bibnamefont{{Karas}}},
  \bibinfo{journal}{\apj} \textbf{\bibinfo{volume}{853}}, \bibinfo{eid}{53}
  (\bibinfo{year}{2018}), \eprint{1801.01576}.

\bibitem[{\citenamefont{{Benavides-Gallego}
  et~al.}(2019)\citenamefont{{Benavides-Gallego}, {Abdujabbarov}, {Malafarina},
  {Ahmedov}, and {Bambi}}}]{2019PhRvD..99d4012B}
\bibinfo{author}{\bibfnamefont{C.~A.} \bibnamefont{{Benavides-Gallego}}},
  \bibinfo{author}{\bibfnamefont{A.}~\bibnamefont{{Abdujabbarov}}},
  \bibinfo{author}{\bibfnamefont{D.}~\bibnamefont{{Malafarina}}},
  \bibinfo{author}{\bibfnamefont{B.}~\bibnamefont{{Ahmedov}}},
  \bibnamefont{and} \bibinfo{author}{\bibfnamefont{C.}~\bibnamefont{{Bambi}}},
  \bibinfo{journal}{\prd} \textbf{\bibinfo{volume}{99}}, \bibinfo{eid}{044012}
  (\bibinfo{year}{2019}), \eprint{1812.04846}.

\bibitem[{\citenamefont{{Stuchl{\'{\i}}k} and {Kolo{\v
  s}}}(2019)}]{Stu-Kol:2019:in_preparation:}
\bibinfo{author}{\bibfnamefont{Z.}~\bibnamefont{{Stuchl{\'{\i}}k}}}
  \bibnamefont{and} \bibinfo{author}{\bibfnamefont{M.}~\bibnamefont{{Kolo{\v
  s}}}} (\bibinfo{year}{2019}), \eprint{in preparation}.

\bibitem[{\citenamefont{{Remillard} and
  {McClintock}}(2006)}]{Rem-McCli:2006:ARAA:}
\bibinfo{author}{\bibfnamefont{R.~A.} \bibnamefont{{Remillard}}}
  \bibnamefont{and} \bibinfo{author}{\bibfnamefont{J.~E.}
  \bibnamefont{{McClintock}}}, \bibinfo{journal}{\araa}
  \textbf{\bibinfo{volume}{44}}, \bibinfo{pages}{49} (\bibinfo{year}{2006}),
  \eprint{astro-ph/0606352}.

\bibitem[{\citenamefont{{Karak} et~al.}(2010)\citenamefont{{Karak}, {Dutta},
  and {Mukhopadhyay}}}]{Kar-Dut-Muk:APJ:2010:}
\bibinfo{author}{\bibfnamefont{B.~B.} \bibnamefont{{Karak}}},
  \bibinfo{author}{\bibfnamefont{J.}~\bibnamefont{{Dutta}}}, \bibnamefont{and}
  \bibinfo{author}{\bibfnamefont{B.}~\bibnamefont{{Mukhopadhyay}}},
  \bibinfo{journal}{\apj} \textbf{\bibinfo{volume}{708}}, \bibinfo{pages}{862}
  (\bibinfo{year}{2010}), \eprint{0911.1701}.

\bibitem[{\citenamefont{{Mannattil} et~al.}(2016)\citenamefont{{Mannattil},
  {Gupta}, and {Chakraborty}}}]{Man-Gup-Cha:APJ:2016:}
\bibinfo{author}{\bibfnamefont{M.}~\bibnamefont{{Mannattil}}},
  \bibinfo{author}{\bibfnamefont{H.}~\bibnamefont{{Gupta}}}, \bibnamefont{and}
  \bibinfo{author}{\bibfnamefont{S.}~\bibnamefont{{Chakraborty}}},
  \bibinfo{journal}{\apj} \textbf{\bibinfo{volume}{833}}, \bibinfo{eid}{208}
  (\bibinfo{year}{2016}), \eprint{1611.02264}.

\bibitem[{\citenamefont{{Sukov{\'a}} et~al.}(2016)\citenamefont{{Sukov{\'a}},
  {Grzedzielski}, and {Janiuk}}}]{Suk-Grz-Jan:AAP:2016:}
\bibinfo{author}{\bibfnamefont{P.}~\bibnamefont{{Sukov{\'a}}}},
  \bibinfo{author}{\bibfnamefont{M.}~\bibnamefont{{Grzedzielski}}},
  \bibnamefont{and} \bibinfo{author}{\bibfnamefont{A.}~\bibnamefont{{Janiuk}}},
  \bibinfo{journal}{\aap} \textbf{\bibinfo{volume}{586}}, \bibinfo{eid}{A143}
  (\bibinfo{year}{2016}), \eprint{1506.02526}.

\bibitem[{\citenamefont{{Huppenkothen}
  et~al.}(2017)\citenamefont{{Huppenkothen}, {Heil}, {Hogg}, and
  {Mueller}}}]{Hup-etal:MNRAS:2017:}
\bibinfo{author}{\bibfnamefont{D.}~\bibnamefont{{Huppenkothen}}},
  \bibinfo{author}{\bibfnamefont{L.~M.} \bibnamefont{{Heil}}},
  \bibinfo{author}{\bibfnamefont{D.~W.} \bibnamefont{{Hogg}}},
  \bibnamefont{and}
  \bibinfo{author}{\bibfnamefont{A.}~\bibnamefont{{Mueller}}},
  \bibinfo{journal}{\mnras} \textbf{\bibinfo{volume}{466}},
  \bibinfo{pages}{2364} (\bibinfo{year}{2017}), \eprint{1611.01332}.

\bibitem[{\citenamefont{{Wald}}(1984)}]{Wald:1984:book:}
\bibinfo{author}{\bibfnamefont{R.~M.} \bibnamefont{{Wald}}},
  \emph{\bibinfo{title}{{General relativity}}} (\bibinfo{publisher}{University
  of Chicago Press, Chicago}, \bibinfo{year}{1984}).

\bibitem[{\citenamefont{{Parthasarathy}
  et~al.}(1986)\citenamefont{{Parthasarathy}, {Wagh}, {Dhurandhar}, and
  {Dadhich}}}]{Par-Wag-Dad:APJ:1986:}
\bibinfo{author}{\bibfnamefont{S.}~\bibnamefont{{Parthasarathy}}},
  \bibinfo{author}{\bibfnamefont{S.~M.} \bibnamefont{{Wagh}}},
  \bibinfo{author}{\bibfnamefont{S.~V.} \bibnamefont{{Dhurandhar}}},
  \bibnamefont{and}
  \bibinfo{author}{\bibfnamefont{N.}~\bibnamefont{{Dadhich}}},
  \bibinfo{journal}{\apj} \textbf{\bibinfo{volume}{307}}, \bibinfo{pages}{38}
  (\bibinfo{year}{1986}).

\bibitem[{\citenamefont{{Dadhich} et~al.}(2018)\citenamefont{{Dadhich},
  {Tursunov}, {Ahmedov}, and {Stuchl{\'{\i}}k}}}]{Dadhich-etal:MNRAS:2018:}
\bibinfo{author}{\bibfnamefont{N.}~\bibnamefont{{Dadhich}}},
  \bibinfo{author}{\bibfnamefont{A.}~\bibnamefont{{Tursunov}}},
  \bibinfo{author}{\bibfnamefont{B.}~\bibnamefont{{Ahmedov}}},
  \bibnamefont{and}
  \bibinfo{author}{\bibfnamefont{Z.}~\bibnamefont{{Stuchl{\'{\i}}k}}},
  \bibinfo{journal}{\mnras} \textbf{\bibinfo{volume}{478}},
  \bibinfo{pages}{L89} (\bibinfo{year}{2018}), \eprint{1804.09679}.

\bibitem[{\citenamefont{{Bardeen} and {Petterson}}(1975)}]{Bar-Pet:1975:ApJ:}
\bibinfo{author}{\bibfnamefont{J.~M.} \bibnamefont{{Bardeen}}}
  \bibnamefont{and} \bibinfo{author}{\bibfnamefont{J.~A.}
  \bibnamefont{{Petterson}}}, \bibinfo{journal}{\apjl}
  \textbf{\bibinfo{volume}{195}}, \bibinfo{pages}{L65} (\bibinfo{year}{1975}).

\bibitem[{\citenamefont{{Stuchl{\'{\i}}k}
  et~al.}(2013)\citenamefont{{Stuchl{\'{\i}}k}, {Kotrlov{\'a}}, and
  {T{\"o}r{\"o}k}}}]{Stu-Kot-Tor:2013:ASTRA:}
\bibinfo{author}{\bibfnamefont{Z.}~\bibnamefont{{Stuchl{\'{\i}}k}}},
  \bibinfo{author}{\bibfnamefont{A.}~\bibnamefont{{Kotrlov{\'a}}}},
  \bibnamefont{and}
  \bibinfo{author}{\bibfnamefont{G.}~\bibnamefont{{T{\"o}r{\"o}k}}},
  \bibinfo{journal}{\aap} \textbf{\bibinfo{volume}{552}}, \bibinfo{eid}{A10}
  (\bibinfo{year}{2013}), \eprint{1305.3552}.

\bibitem[{\citenamefont{{Al Zahrani}}(2014)}]{Zah:PRD:2014:}
\bibinfo{author}{\bibfnamefont{A.~M.} \bibnamefont{{Al Zahrani}}},
  \bibinfo{journal}{\prd} \textbf{\bibinfo{volume}{90}}, \bibinfo{eid}{044012}
  (\bibinfo{year}{2014}), \eprint{1407.7069}.

\bibitem[{\citenamefont{{Tursunov}
  et~al.}(2018{\natexlab{a}})}]{Tur-etal:2018:submitted}
\bibinfo{author}{\bibfnamefont{A.}~\bibnamefont{{Tursunov}}}
  \bibnamefont{et~al.}, \bibinfo{journal}{(submitted)}
  (\bibinfo{year}{2018}{\natexlab{a}}).

\bibitem[{\citenamefont{{Tursunov}
  et~al.}(2018{\natexlab{b}})\citenamefont{{Tursunov}, {Kolo{\v s}},
  {Stuchl{\'{\i}}k}, and {Galtsov}}}]{Tur-etal:APJ:2018:}
\bibinfo{author}{\bibfnamefont{A.}~\bibnamefont{{Tursunov}}},
  \bibinfo{author}{\bibfnamefont{M.}~\bibnamefont{{Kolo{\v s}}}},
  \bibinfo{author}{\bibfnamefont{Z.}~\bibnamefont{{Stuchl{\'{\i}}k}}},
  \bibnamefont{and} \bibinfo{author}{\bibfnamefont{D.~V.}
  \bibnamefont{{Galtsov}}}, \bibinfo{journal}{\apj}
  \textbf{\bibinfo{volume}{861}}, \bibinfo{eid}{2}
  (\bibinfo{year}{2018}{\natexlab{b}}), \eprint{1803.09682}.

\bibitem[{\citenamefont{Schroeder}(1991)}]{1991fcpl.book.....S}
\bibinfo{author}{\bibfnamefont{M.}~\bibnamefont{Schroeder}},
  \emph{\bibinfo{title}{Fractals, Chaos, Power Laws: Minutes From an Infinite
  Paradise}}, vol.~\bibinfo{volume}{44} (\bibinfo{year}{1991}).

\bibitem[{\citenamefont{{Grassberger} and
  {Procaccia}}(1983)}]{1983PhRvL..50..346G}
\bibinfo{author}{\bibfnamefont{P.}~\bibnamefont{{Grassberger}}}
  \bibnamefont{and}
  \bibinfo{author}{\bibfnamefont{I.}~\bibnamefont{{Procaccia}}},
  \bibinfo{journal}{Physical Review Letters} \textbf{\bibinfo{volume}{50}},
  \bibinfo{pages}{346} (\bibinfo{year}{1983}).

\bibitem[{\citenamefont{{Kantz}}(1994)}]{1994PhLA..185...77K}
\bibinfo{author}{\bibfnamefont{H.}~\bibnamefont{{Kantz}}},
  \bibinfo{journal}{Physics Letters A} \textbf{\bibinfo{volume}{185}},
  \bibinfo{pages}{77} (\bibinfo{year}{1994}).

\bibitem[{\citenamefont{{Zbilut} and {Webber}}(1992)}]{1992PhLA..171..199Z}
\bibinfo{author}{\bibfnamefont{J.~P.} \bibnamefont{{Zbilut}}} \bibnamefont{and}
  \bibinfo{author}{\bibfnamefont{C.~L.} \bibnamefont{{Webber}}},
  \bibinfo{journal}{Physics Letters A} \textbf{\bibinfo{volume}{171}},
  \bibinfo{pages}{199} (\bibinfo{year}{1992}).

\bibitem[{\citenamefont{{Marwan}}(2008)}]{2008EPJST.164....3M}
\bibinfo{author}{\bibfnamefont{N.}~\bibnamefont{{Marwan}}},
  \bibinfo{journal}{European Physical Journal Special Topics}
  \textbf{\bibinfo{volume}{164}}, \bibinfo{pages}{3} (\bibinfo{year}{2008}),
  \eprint{1709.09971}.

\bibitem[{\citenamefont{{Eckmann} et~al.}(1987)\citenamefont{{Eckmann},
  {Oliffson Kamphorst}, and {Ruelle}}}]{1987EL......4..973E}
\bibinfo{author}{\bibfnamefont{J.-P.} \bibnamefont{{Eckmann}}},
  \bibinfo{author}{\bibfnamefont{S.}~\bibnamefont{{Oliffson Kamphorst}}},
  \bibnamefont{and} \bibinfo{author}{\bibfnamefont{D.}~\bibnamefont{{Ruelle}}},
  \bibinfo{journal}{EPL (Europhysics Letters)} \textbf{\bibinfo{volume}{4}},
  \bibinfo{pages}{973} (\bibinfo{year}{1987}).

\bibitem[{\citenamefont{{Silver}
  et~al.}(2017{\natexlab{a}})\citenamefont{{Silver}, {Schrittwieser},
  {Simonyan}, {Antonoglou}, {Huang}, {Guez}, {Hubert}, {Baker}, {Lai}, {Bolton}
  et~al.}}]{2017Natur.550..354S}
\bibinfo{author}{\bibfnamefont{D.}~\bibnamefont{{Silver}}},
  \bibinfo{author}{\bibfnamefont{J.}~\bibnamefont{{Schrittwieser}}},
  \bibinfo{author}{\bibfnamefont{K.}~\bibnamefont{{Simonyan}}},
  \bibinfo{author}{\bibfnamefont{I.}~\bibnamefont{{Antonoglou}}},
  \bibinfo{author}{\bibfnamefont{A.}~\bibnamefont{{Huang}}},
  \bibinfo{author}{\bibfnamefont{A.}~\bibnamefont{{Guez}}},
  \bibinfo{author}{\bibfnamefont{T.}~\bibnamefont{{Hubert}}},
  \bibinfo{author}{\bibfnamefont{L.}~\bibnamefont{{Baker}}},
  \bibinfo{author}{\bibfnamefont{M.}~\bibnamefont{{Lai}}},
  \bibinfo{author}{\bibfnamefont{A.}~\bibnamefont{{Bolton}}},
  \bibnamefont{et~al.}, \bibinfo{journal}{\nat} \textbf{\bibinfo{volume}{550}},
  \bibinfo{pages}{354} (\bibinfo{year}{2017}{\natexlab{a}}).

\bibitem[{\citenamefont{{Silver}
  et~al.}(2017{\natexlab{b}})\citenamefont{{Silver}, {Hubert}, {Schrittwieser},
  {Antonoglou}, {Lai}, {Guez}, {Lanctot}, {Sifre}, {Kumaran}, {Graepel}
  et~al.}}]{2017arXiv171201815S}
\bibinfo{author}{\bibfnamefont{D.}~\bibnamefont{{Silver}}},
  \bibinfo{author}{\bibfnamefont{T.}~\bibnamefont{{Hubert}}},
  \bibinfo{author}{\bibfnamefont{J.}~\bibnamefont{{Schrittwieser}}},
  \bibinfo{author}{\bibfnamefont{I.}~\bibnamefont{{Antonoglou}}},
  \bibinfo{author}{\bibfnamefont{M.}~\bibnamefont{{Lai}}},
  \bibinfo{author}{\bibfnamefont{A.}~\bibnamefont{{Guez}}},
  \bibinfo{author}{\bibfnamefont{M.}~\bibnamefont{{Lanctot}}},
  \bibinfo{author}{\bibfnamefont{L.}~\bibnamefont{{Sifre}}},
  \bibinfo{author}{\bibfnamefont{D.}~\bibnamefont{{Kumaran}}},
  \bibinfo{author}{\bibfnamefont{T.}~\bibnamefont{{Graepel}}},
  \bibnamefont{et~al.}, \bibinfo{journal}{ArXiv e-prints}
  (\bibinfo{year}{2017}{\natexlab{b}}), \eprint{1712.01815}.

\bibitem[{\citenamefont{Wolfram|Alpha}(2018{\natexlab{a}})}]{M1}
\bibinfo{author}{\bibnamefont{Wolfram|Alpha}}, \emph{\bibinfo{title}{{Wolfram
  Alpha LLC} "decisiontree" (machine learning method)}}
  (\bibinfo{year}{2018}{\natexlab{a}}),
  \urlprefix\url{https://reference.wolfram.com/language/ref/method/DecisionTree.html}.

\bibitem[{\citenamefont{Wolfram|Alpha}(2018{\natexlab{b}})}]{M2}
\bibinfo{author}{\bibnamefont{Wolfram|Alpha}}, \emph{\bibinfo{title}{{Wolfram
  Alpha LLC} "neuralnetwork" (machine learning method)}}
  (\bibinfo{year}{2018}{\natexlab{b}}),
  \urlprefix\url{https://reference.wolfram.com/language/ref/method/NeuralNetwork.html}.

\bibitem[{\citenamefont{Wolfram|Alpha}(2018{\natexlab{c}})}]{M3}
\bibinfo{author}{\bibnamefont{Wolfram|Alpha}}, \emph{\bibinfo{title}{{Wolfram
  Alpha LLC} "linearregression" (machine learning method)}}
  (\bibinfo{year}{2018}{\natexlab{c}}),
  \urlprefix\url{https://reference.wolfram.com/language/ref/method/LinearRegression.html}.

\bibitem[{\citenamefont{Wolfram|Alpha}(2018{\natexlab{d}})}]{M4}
\bibinfo{author}{\bibnamefont{Wolfram|Alpha}}, \emph{\bibinfo{title}{{Wolfram
  Alpha LLC} "randomforest" (machine learning method)}}
  (\bibinfo{year}{2018}{\natexlab{d}}),
  \urlprefix\url{https://reference.wolfram.com/language/ref/method/RandomForest.html}.

\bibitem[{\citenamefont{Wolfram|Alpha}(2018{\natexlab{e}})}]{M5}
\bibinfo{author}{\bibnamefont{Wolfram|Alpha}}, \emph{\bibinfo{title}{{Wolfram
  Alpha LLC} "gaussianprocess" (machine learning method)}}
  (\bibinfo{year}{2018}{\natexlab{e}}),
  \urlprefix\url{https://reference.wolfram.com/language/ref/method/GaussianProcess.html}.

\bibitem[{\citenamefont{Wolfram|Alpha}(2018{\natexlab{f}})}]{M6}
\bibinfo{author}{\bibnamefont{Wolfram|Alpha}}, \emph{\bibinfo{title}{{Wolfram
  Alpha LLC} "nearestneighbors" (machine learning method)}}
  (\bibinfo{year}{2018}{\natexlab{f}}),
  \urlprefix\url{https://reference.wolfram.com/language/ref/method/NearestNeighbors.html}.

\bibitem[{\citenamefont{Wolfram|Alpha}(2018{\natexlab{g}})}]{M7}
\bibinfo{author}{\bibnamefont{Wolfram|Alpha}}, \emph{\bibinfo{title}{{Wolfram
  Alpha LLC} "gradientboostedtrees" (machine learning method)}}
  (\bibinfo{year}{2018}{\natexlab{g}}),
  \urlprefix\url{https://reference.wolfram.com/language/ref/method/GradientBoostedTrees.html}.

\bibitem[{\citenamefont{{Ott}}(1993)}]{Ott:1993:book:}
\bibinfo{author}{\bibfnamefont{E.}~\bibnamefont{{Ott}}},
  \emph{\bibinfo{title}{{Chaos in dynamical systems}}}
  (\bibinfo{publisher}{Cambridge University Press}, \bibinfo{year}{1993}).

\bibitem[{\citenamefont{Pánis}(2019)}]{Git}
\bibinfo{author}{\bibfnamefont{R.}~\bibnamefont{Pánis}},
  \emph{\bibinfo{title}{Chaos detection}},
  \bibinfo{howpublished}{\url{https://github.com/radim525/Chaos-detection}}
  (\bibinfo{year}{2019}).

\bibitem[{\citenamefont{{Cardoso} et~al.}(2009)\citenamefont{{Cardoso},
  {Miranda}, {Berti}, {Witek}, and {Zanchin}}}]{2009PhRvD..79f4016C}
\bibinfo{author}{\bibfnamefont{V.}~\bibnamefont{{Cardoso}}},
  \bibinfo{author}{\bibfnamefont{A.~S.} \bibnamefont{{Miranda}}},
  \bibinfo{author}{\bibfnamefont{E.}~\bibnamefont{{Berti}}},
  \bibinfo{author}{\bibfnamefont{H.}~\bibnamefont{{Witek}}}, \bibnamefont{and}
  \bibinfo{author}{\bibfnamefont{V.~T.} \bibnamefont{{Zanchin}}},
  \bibinfo{journal}{\prd} \textbf{\bibinfo{volume}{79}}, \bibinfo{eid}{064016}
  (\bibinfo{year}{2009}), \eprint{0812.1806}.

\bibitem[{\citenamefont{{Konoplya} and
  {Stuchl{\'{\i}}k}}(2017)}]{2017PhLB..771..597K}
\bibinfo{author}{\bibfnamefont{R.~A.} \bibnamefont{{Konoplya}}}
  \bibnamefont{and}
  \bibinfo{author}{\bibfnamefont{Z.}~\bibnamefont{{Stuchl{\'{\i}}k}}},
  \bibinfo{journal}{Physics Letters B} \textbf{\bibinfo{volume}{771}},
  \bibinfo{pages}{597} (\bibinfo{year}{2017}), \eprint{1705.05928}.

\bibitem[{\citenamefont{{Toshmatov} et~al.}(2018)\citenamefont{{Toshmatov},
  {Stuchl{\'{\i}}k}, {Schee}, and {Ahmedov}}}]{2018PhRvD..97h4058T}
\bibinfo{author}{\bibfnamefont{B.}~\bibnamefont{{Toshmatov}}},
  \bibinfo{author}{\bibfnamefont{Z.}~\bibnamefont{{Stuchl{\'{\i}}k}}},
  \bibinfo{author}{\bibfnamefont{J.}~\bibnamefont{{Schee}}}, \bibnamefont{and}
  \bibinfo{author}{\bibfnamefont{B.}~\bibnamefont{{Ahmedov}}},
  \bibinfo{journal}{\prd} \textbf{\bibinfo{volume}{97}}, \bibinfo{eid}{084058}
  (\bibinfo{year}{2018}), \eprint{1805.00240}.

\end{thebibliography}

\end{document}